\pdfoutput=1
\RequirePackage[2020-02-02]{latexrelease}
\documentclass[prd,reprint,twocolumn,superscriptaddress,letterpaper,nofootinbib,
               nopreprintnumbers,showpacs]{revtex4-1}

\usepackage{amsmath}
\usepackage{microtype}
\usepackage{siunitx}
\usepackage{txfonts}
\usepackage{verbatim}
\usepackage{dcolumn}

\usepackage[usenames, dvipsnames, svgnames, table]{xcolor}
\usepackage[colorlinks=true, citecolor=RoyalBlue,
            linkcolor=BrickRed, urlcolor=ForestGreen]{hyperref}
\usepackage{graphicx}
\usepackage[justification=justified]{caption}
\usepackage[justification=justified]{subcaption}
\usepackage{float}
\usepackage{lineno}
\usepackage{multirow}
\usepackage{placeins}

\usepackage{acronym}
\usepackage{ifthen}
\usepackage{svn-multi}

\newcommand*\patchAmsMathEnvironmentForLineno[1]{%
  \expandafter\let\csname old#1\expandafter\endcsname\csname #1\endcsname
  \expandafter\let\csname oldend#1\expandafter\endcsname\csname end#1\endcsname
  \renewenvironment{#1}%
     {\linenomath\csname old#1\endcsname}%
     {\csname oldend#1\endcsname\endlinenomath}}%
\newcommand*\patchBothAmsMathEnvironmentsForLineno[1]{%
  \patchAmsMathEnvironmentForLineno{#1}%
  \patchAmsMathEnvironmentForLineno{#1*}}%
\AtBeginDocument{%
\patchBothAmsMathEnvironmentsForLineno{equation}%
\patchBothAmsMathEnvironmentsForLineno{align}%
\patchBothAmsMathEnvironmentsForLineno{flalign}%
\patchBothAmsMathEnvironmentsForLineno{alignat}%
\patchBothAmsMathEnvironmentsForLineno{gather}%
\patchBothAmsMathEnvironmentsForLineno{multline}%
}

\begin{document}

\title{Improving LIGO calibration accuracy by using time-dependent filters to compensate for temporal variations}
\author{M. Wade}
\affiliation{Kenyon College, Gambier, OH 43022, USA}
\author{A. D. Viets}
\affiliation{Concordia University Wisconsin, Mequon, WI 53097, USA}
\author{T. Chmiel}
\affiliation{Kenyon College, Gambier, OH 43022, USA}
\affiliation{University of Chicago, Chicago, IL 60637, USA}
\author{M. Stover}
\affiliation{Kenyon College, Gambier, OH 43022, USA}
\affiliation{University of Illinois Urbana-Champaign, Urbana, IL 61801, USA}
\author{L. Wade}
\affiliation{Kenyon College, Gambier, OH 43022, USA}

\begin{abstract}
The response of the Advanced LIGO interferometers is known to vary with time \cite{Darkhan}.  Accurate calibration of the interferometers must therefore track and compensate for temporal variations in calibration model parameters.  These variations were tracked during the first three Advanced LIGO observing runs, and compensation for some of them has been implemented in the calibration procedure.  During the second observing run, multiplicative corrections to the interferometer response were applied while producing calibrated strain data both in real time and in high latency.  In a high-latency calibration produced after the second observing run and during the entirety of the third observing run, a correction requiring periodic filter updates was applied to the calibration--the time dependence of the coupled cavity pole frequency $f_{\rm cc}$.  This paper describes the methods developed to compensate for variations in the interferometer response requiring time-dependent filters, including variable zeros, poles, gains, and time delays.  The described methods were used to provide compensation for well-modeled time dependence of the interferometer response, which has helped to reduce systematic errors in the calibration to $<2$\% in magnitude and $<2^{\circ}$ in phase across LIGO's most sensitive frequency band of 20 - 2000 Hz \cite{Sun:2020wke,Sun:2021qcg}.  Additionally, this paper shows how such compensation is relevant for astrophysical inference studies by reducing uncertainty and bias in the sky localization for a simulated binary neutron star merger.
\end{abstract}

\maketitle

\section{Introduction}
\label{sec:intro}

The Advanced Laser Interferometer Gravitational-wave Observatory (Advanced LIGO) \cite{TheLIGOScientific:2014jea} is a network of two ground-based gravitational-wave detectors, located in Hanford, WA (H1), and in Livingston, LA (L1).  
To date, Advanced LIGO has completed three observing runs (O1, O2, and O3) and, together with the Advanced Virgo detector \cite{VIRGO:2014yos}, made numerous detections of gravitational waves (GWs) originating from transient astrophysical sources \cite{LIGOScientific:2018mvr, PhysRevX.11.021053,LIGOScientific:2021usb,LIGOScientific:2021djp}.  
The first step in analyzing the data from these ground-based gravitational-wave interferometers is the reconstruction of the interferometer strain time series.  
This process, known as calibration, involves developing physically motivated models of the interferometer optics and feedback systems in order to compute the incident detector strain from the interferometer's digital feedback loops.  
These developed models, referred to below as the calibration models,  involve parameters that vary in time in ways that reflect the evolution of physical systems within the interferometer.  

During all three Advanced LIGO observing runs the calibration process has compensated for temporal variations in the calibration model that can be applied as multiplicative correction factors, which we will refer to as time-dependent correction factor (TDCF) multipliers.  
The methods for such compensations are discussed in detail in Ref.~\cite{Darkhan, hoft}.  
However, the calibrated data from O1 and the low-latency calibrated data from O2 left parametric updates requiring the generation of new calibration filters, which we will refer to as TDCF filters, uncompensated.  
In this work, we describe the methodology used to incorporate TDCF filters in the production of the high-latency calibrated data in O2 and throughout O3.
This same methodology is anticipated to be used in the upcoming O4 observing run as well.
Additionally, we investigate how compensating for TDCF filters could impact our parameter inference with a simulated binary neutron star signal.

\begin{figure}[!b]
    \centering
    \includegraphics[width=\columnwidth]{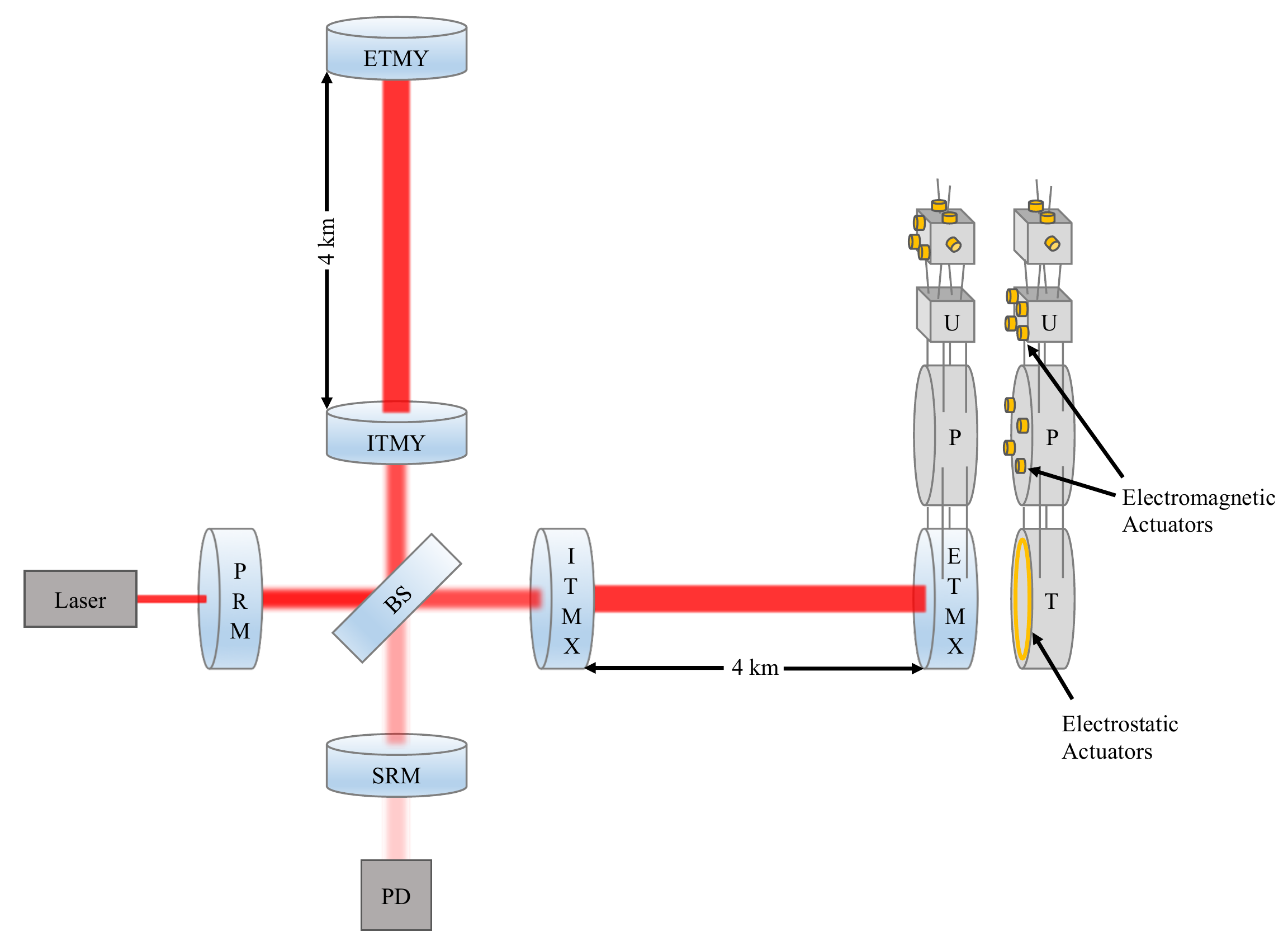}
    \caption{Simplified diagram of an Advanced LIGO detector. Laser light is sent through a power recycling mirror (PRM) and is split by a beamsplitter (BS) to enter a pair of Fabry-P\'erot cavities.  Light is held inside the cavities by mirrors on the input test masses (ITMX and ITMY) and the end test masses (ETMX and ETMY).  After exiting the cavities, the light recombines nearly 180$^{\circ}$ out of phase at the beamsplitter.  Light that passes through the signal recycling mirror (SRM) is sent to a photodetector (PD).  One of the dual-chain quadruple pendulum suspension systems with actuators is shown on the right.  Differential arm motion is actively suppressed in the lowest three stages, called the test mass (T) stage, the penultimate (P) stage, and the upper-intermediate (U) stage.
\label{fig:aLIGO}}
\end{figure}

Each Advanced LIGO detector consists of two orthogonal 4-km arms called the X arm and the Y arm. 
Near-infrared (1064-nm) laser light is passed through a beamsplitter into each arm, before being reflected back to the beamsplitter by mirrors on the end test masses (ETMs). 
In an unperturbed state, the length of each arm is held such that a weak light signal will exit the detector at the GW readout port (see Fig.~\ref{fig:aLIGO}) \cite{Fricke_2012}.
Gravitational waves incident on the LIGO interferometers cause fluctuations in this readout by inducing changes in the differential arm length (DARM) degree of freedom in the detectors:
\begin{equation}
\Delta L_{\rm free}(t) = \Delta L_{\rm X}(t) - \Delta L_{\rm Y}(t).
\end{equation}
Changes in DARM cause fluctuations in the intensity of the laser light at the GW readout port, which is recorded as a 16384~Hz digital error signal $d_{\rm err}$ in arbitrary units called counts. 
In order to improve sensitivity by increasing the power stored in the arms, additional mirrors are placed in the arms near the beamsplitter, forming a 4-km Fabry-P\'erot cavity in each arm. 
A power-recycling mirror is also included before the beamsplitter to increase the power entering the detector, further improving the sensitivity. 
Lastly, in order to enhance sensitivity in LIGO's most sensitive frequency band [20 Hz, 2 kHz], a signal-recycling mirror is placed just before the GW readout port.

\begin{figure*}[!t]
    \centering
    \includegraphics[width=0.9\textwidth]{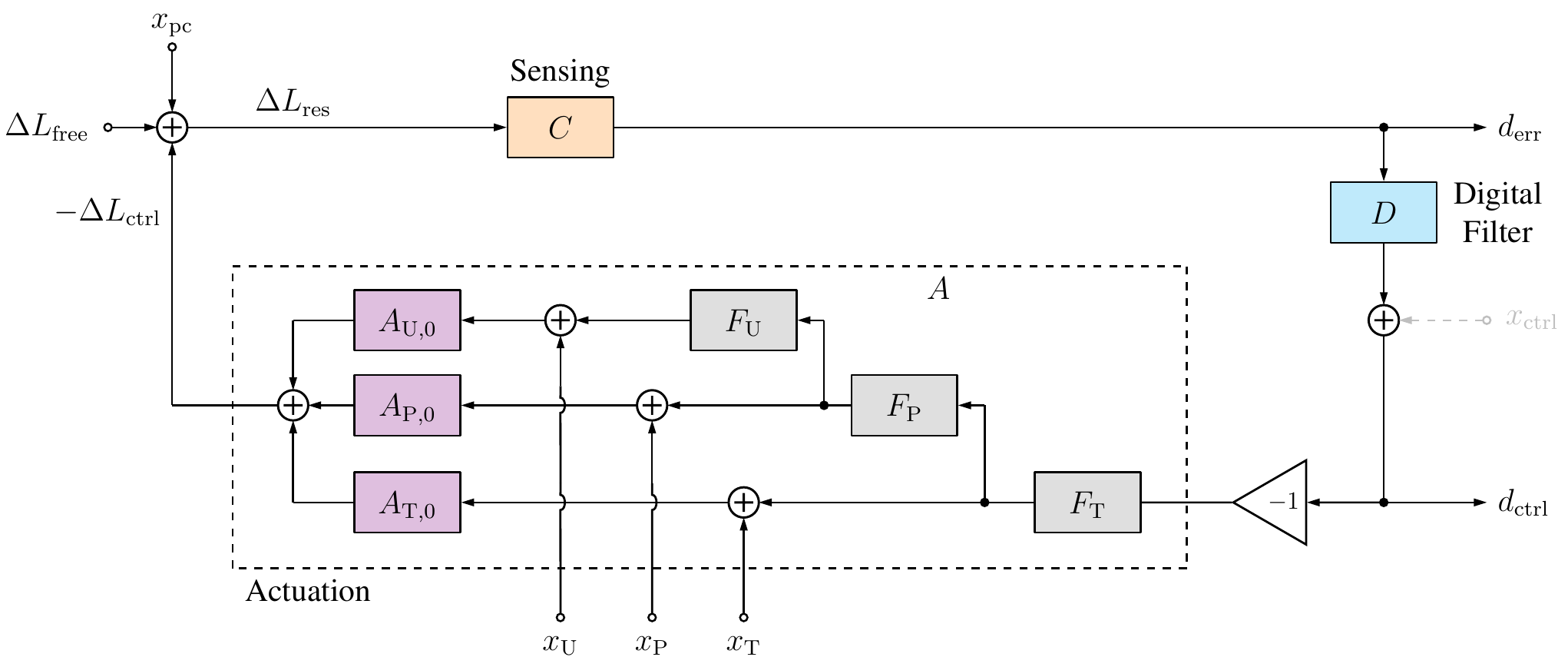}
    \caption{Block diagram of the Advanced LIGO differential arm (DARM) length feedback control loop during the third observing run (O3).
The sensing function $C$ represents the detector's response to residual DARM motion in the test masses.
The digital filter $D$ converts the error signal $d_{\rm err}$ to the control signal $d_{\rm ctrl}$.
The actuation function $A$ is split into three stages, represented by $A_{\rm U}$, $A_{\rm P}$, and $A_{\rm T}$, corresponding to the three suspension stages at which DARM motion is actively suppressed.
External DARM motion $\Delta L_{\rm free}$ and the controlled length differential $\Delta L_{\rm ctrl}$ enter the diagram in the upper-left corner to produce the residual DARM motion $\Delta L_{\rm res}$ sensed by the detector.
Known excitations are intentionally injected into the DARM loop using both the actuation system (through the injections $x_{\rm T}$, $x_{\rm P}$, and $x_{\rm U}$) and a radiation pressure actuator known as a photon calibrator ($x_{\rm pc}$).  The transfer functions $F_{\rm T}$, $F_{\rm P}$, and $F_{\rm U}$ are filter functions that occur before the injections are added.  During Advanced LIGO's second observing run (O2), the injection $x_{\rm ctrl}$ was also added to $d_{\rm ctrl}$. \label{fig:DARM_loop}}
\end{figure*}

Despite the use of seismic isolation and the quadruple suspension systems to attenuate excess noise at low frequencies, the detectors cannot achieve a resonant low-noise state without additional mitigation of noise.  
Therefore, the error signal is filtered with a digital filter $D$ to produce a control signal $d_{\rm ctrl} = D \ast d_{\rm err}$, where $\ast$ denotes a time-domain convolution, equivalent to a frequency-domain multiplication.  
The control signal is fed into a set of actuators that form a quadruple pendulum similar to that of the test masses, separated by 0.5 cm from the test masses (Fig.~\ref{fig:aLIGO}) \cite{Robertson:2002jf, Aston:2012ona}.  

Although the actuators on both the X and Y arms are used to control the common arm length degree of freedom, the DARM feedback is sent to only one set of actuators.  
The digital control signal $d_{\rm ctrl}$ is related to the analog controlled length differential $\Delta L_{\rm ctrl}$ through the actuation function $A$: $\Delta L_{\rm ctrl} = A \ast d_{\rm ctrl}$.  
The actuation system removes the controlled length differential $\Delta L_{\rm ctrl}$ from $\Delta L_{\rm free}$ to form the residual length differential sensed by the detector, 
\begin{equation}
\Delta L_{\rm res} = \Delta L_{\rm free} - \Delta L_{\rm ctrl} \ .
\end{equation}
The digital error signal $d_{\rm err}$ is related to the residual length differential by a sensing function $C$ defined by $d_{\rm err} = C \ast \Delta L_{\rm res}$.  
This feedback loop, known as the DARM loop, is diagrammed in Fig.~\ref{fig:DARM_loop}.  
The DARM loop can be used to solve for $\Delta L_{\rm free}$, the result being
\begin{equation}\label{eq:DeltaL}
\Delta L_{\rm free} = C^{-1} \ast d_{\rm err} + A \ast d_{\rm ctrl},
\end{equation}
or expressed alternatively,
\begin{equation}\label{eq:DeltaLwithR}
\Delta L_{\rm free} = R \ast d_{\rm err},
\end{equation}
where $\tilde{R}(f) = [1 + \tilde{A}(f) \tilde{D}(f) \tilde{C}(f)] / \tilde{C}(f)$ is the response function of the interferometer, represented here in the frequency domain for simplicity.
The final product of Advanced LIGO's calibration process is the dimensionless strain,
\begin{equation}
h(t) = \frac{\Delta L_{\rm free}(t)}{L},
\end{equation}
where $L = (L_{\rm X} + L_{\rm Y}) / 2$ is the average arm length.


Static reference models for $A$ and $C$ are produced in the frequency domain at the beginning of observing runs~\cite{calcompanion}, and periodically throughout observing runs, to check calibration accuracy and inform a calibration uncertainty estimate, described in other works~\cite{Sun:2021qcg,Sun:2020wke,Craig}.  
These models are used to produce time-domain filters for $A$ and $C^{-1}$ which are then implemented into LIGO's calibration pipelines~\cite{hoft}.  
For more detailed information on the methods used to compute the calibrated strain $h(t)$, see Refs.~\cite{hoft, VietsDissertation}.

Measurements used to determine the parameters in the reference models for the actuation and sensing functions (see Sec.~\ref{sec:CalibrationModels}) are taken using an auxiliary laser, known as a photon calibrator (Pcal) \cite{pcal}.  
There are two Pcals, one located at each end test mass (ETM).  
Radiation pressure exerted by a Pcal on an ETM produces a fiducial displacement that acts as the primary reference for absolute displacement calibration.  
The overall 1-$\sigma$ uncertainty in the displacements induced by the Pcals is 0.41\% during O3 \cite{O3pcal}. 

Swept-sine injections, which are sinusoidal excitations where the frequency is swept across the band at set intervals, are made using a Pcal and the actuation system, and these injections are used to find best-fit values for the parameters of $A$ and $C$.   
In addition, sinusoidal excitations at select frequencies, called calibration lines, are continuously injected using a Pcal and the actuation system, in order to track parameters that are known to show slow temporal variations \cite{Darkhan}.  
Occasionally, broadband injections, which are excitations that span a broad range of frequencies, are also made using a Pcal to check calibration accuracy across all frequencies.  
Broadband injections and swept sine injections only occur when the interferometer is not being actively used for astrophysical observation.  
The calibration lines are run constantly throughout the observing run in order to track calibration parameters throughout data being used for astrophysical observation.

The rest of this paper is structured as follows.  In Sec.~\ref{sec:CalibrationModels} we give a brief introduction to the fundamental models used in the calibration procedure.  
In Sec.~\ref{sec:kappas}, we outline the methods used to compute time-dependent correction factors (TDCFs) for the calibration models.  
Sec.~\ref{sec:filterupdates} describes the methods developed for computing and applying TDCF filters.  
Sec.~\ref{sec:calibrationAccuracy} discusses the impact of applying TDCF filters on calibration accuracy.  
Sec.~\ref{sec:astrophysics} discusses the impact of some example systematic calibration errors arising from not compensating for temporal variations in the calibration models on an astrophysical analysis.  
Finally, we conclude in Sec.~\ref{sec:conclusion}.

\section{Calibration Models}
\label{sec:CalibrationModels}

As seen in Fig.~\ref{fig:DARM_loop} and Eq.~\eqref{eq:DeltaL}, the sensing and actuation functions are the key components of the interferometer response needed to reconstruct the dimensionless strain from the digital control loop outputs, $d_{\rm err}$ and $d_{\rm ctrl}$.
The sensing function models the interferometer's response to residual differential arm motion in the ETMs.  The actuation function models the active suppression of external DARM motion \cite{calcompanion}.

The full model for the sensing function is
\begin{eqnarray}\label{eq:C}
\tilde{C}(f; t) &=& \kappa_{\rm C}(t)  \left(\frac{H_{\rm C}}{1 + if/f_{\rm cc}(t)}\right) \\
\nonumber
&& \times \left(\frac{f^2}{f^2 + f_{\rm s}(t)^2 - i f f_{\rm s}(t) / Q(t)}\right)  \\
\nonumber
&& \times \ C_{\rm R}(f) \, \rm{exp}\left[{-2\pi \it{if}\tau_{\rm C}}\right] \ .
\end{eqnarray}
The gain $H_{\rm C}$ represents the conversion from meters of DARM displacement to counts as measured in the reference model.  
The dimensionless scalar $\kappa_{\rm C}(t)$ has a nominal value of one and encodes the time-dependence of the gain $H_{\rm C}$, observed to fluctuate by $\sim$10\%.  
The coupled cavity pole frequency $f_{\rm cc}$ is the characteristic frequency at which the detector response is significantly attenuated due to finite average photon storage time in the Fabry-P\'erot cavities.  
During the second part of the third observing run (O3b), the coupled cavity pole frequency had a nominal value of 411 Hz at H1 and 461 Hz at L1, and it was observed to fluctuate as much as $\sim$20 Hz.  
$\tau_{\rm C}$ is a constant time delay due to light-travel time across the length of each arm and an additional time delay in acquiring the digital signal.  
The factor $C_{\rm R}(f)$ encodes the remaining frequency dependence above $\sim$1 kHz due to photodiode electronics and signal-processing filters.

The second term in parenthesis represents the impact of the slightly detuned signal recycling cavity (SRC) on the sensing function, impactful mainly below $\sim$50 Hz.  $f_{\rm s}(t)$ and $Q(t)$ are, respectively, the resonant frequency and quality factor of the optical anti-spring of the SRC~\cite{Craig, PhysRevD.74.022001, hall2017long}.  
An optical spring (or anti-spring) exists in an optomechanical cavity if there is a linear relationship between the length of the cavity and the radiation pressure on the mirrors.  
By design, $f_{\rm s}$ should remain close to zero and the SRC should have no impact on the frequency-dependence of the sensing function.  Nonzero values of $f_{\rm s}$ indicate detuning of the SRC.
During O2, the SRC at H1 was found to have an optical anti-spring with a time-varying resonant frequency $f_{\rm s} \lessapprox 10$ Hz.
During O3, H1 and L1 observed measurements indicative of both optical spring and anti-spring behavior in the SRC.  
Moreover, at H1, there is evidence of a two-way cross-coupling of the DARM feedback control loop with feedback used to control angular motion of the test masses (L2A2L cross-coupling)~\cite{Sun:2020wke, Brooks2021}.
This was likely due to the off-center positioning of the laser beam necessary to avoid a point defect in the test mass.
Although a mathematical model is not yet available for L2A2L cross-coupling, it is also believed to impact the sensing function at low frequencies.  
It is therefore unclear how much impact SRC detuning and L2A2L cross-coupling have individually on the low-frequency sensing function, and whether or not each impact is time-dependent.
It is clear that at least one of these low-frequency effects is time-dependent, since the sensing function undergoes a significant, measurable change at low frequencies during thermalization in the first $\sim$hour of low-noise operation.
Although this change was uncompensated, it impacted only a small portion of total observing time and did not dominate the calibration error estimate.  The increase in systematic calibration error was accounted for in cases of events detected by the H1 detector during thermalization \cite{Sun:2020wke}.

The full actuation model we will use for this analysis is given by
\begin{eqnarray}\label{eq:A}
\tilde{A}(f;t) &=& \Big[ \kappa_{\rm T}(t) e^{2\pi i f \tau_{\rm T}(t)} \tilde{A}_{\rm T}(f) \\
&& + \kappa_{\rm P}(t) e^{2\pi i f \tau_{\rm P}(t)} \tilde{A}_{\rm P}(f) \nonumber \\
&& + \kappa_{\rm U}(t) e^{2\pi i f \tau_{\rm U}(t)} \tilde{A}_{\rm U}(f) \Big] \exp\left[-2\pi if\tau_{\rm A}\right] \nonumber ,
\end{eqnarray}
where $\tilde{A}_i(f) = \tilde{A}_{i,0}(f) {\displaystyle \prod_{j \leq i}} \tilde{F}_{j}(f)$ represents the frequency response of the $i$-th actuator for $i \in \{{\rm T, P, U}\}$.
Lower-frequency content of $d_{\rm ctrl}$ is directed to higher stages of actuation, and higher-frequency content is directed to lower stages.  
$\tau_{\rm A}$ is a constant computational time delay.
$\kappa_{\rm T}(t)$, $\kappa_{\rm P}(t)$, and $\kappa_{\rm U}(t)$, all nominally equal to one, represent the time dependence of the strength of each stage of actuation.
$\tau_{\rm T}$, $\tau_{\rm P}$, and $\tau_{\rm U}$, all nominally zero, represent time-dependent time advances relative to the timing of the reference model for each $\tilde{A}_i$.  
Before O3, the time dependence of the penultimate and upper-intermediate stages of actuation were tracked together using the factors $\kappa_{\rm PU}(t)$ and $\tau_{\rm PU}(t)$.  
This was tracked using a calibration line injected via $x_{\rm ctrl}$ into $d_{\rm ctrl}$ instead of the penultimate and upper-intermediate stages of actuation (see Fig.~\ref{fig:DARM_loop}).

$\kappa_{\rm T}$ has been observed to fluctuate by $\sim$10\%, and $\kappa_{\rm P}$ and $\kappa_{\rm U}$ have been observed to fluctuate by $\sim$5\%.  
The variable time delays $\tau_i$ have nominal values of zero and are generally expected to remain small, but as suggested in Sec.~\ref{sec:calibrationAccuracy}, may drift as much as $\sim$\SI{100}{\micro\second}\footnote{Previous publications treat $\kappa_i$ as complex parameters whose imaginary parts are expected to be close to zero instead of defining $\tau_i$.}.  
The most likely source of true changes in these time advances is variation in computational time delays in the digital portion of the actuation function, observed as occasional sudden changes in the values of the $\tau_i$.  
However, the breakdown of the approximation used to estimate the $\kappa_i$, described in the next section, can also lead to erroneous changes in the estimates of the $\tau_i$, making it impratical to compensate for their time dependence (see Appendix~\ref{app:tau}).


Before the end of O2, the reconstruction of the calibrated strain $h(t)$ included only compensation for the time dependence of $\kappa_{\rm T}$, $\kappa_{\rm PU}$, and $\kappa_{\rm C}$.  
These corrections can be applied to the sensing and actuation functions as multiplicative factors, and are therefore referred to as TDCF multipliers.
It is now possible to compensate for all modeled time dependence, including time dependence requiring updates to time-domain filters constructed from the calibration models, using the adaptive filtering techniques described in Sec.~\ref{sec:filterupdates}.

\section{Computing Time-dependent Correction Factors}
\label{sec:kappas}

In order to measure changes in the time-dependent correction factors (TDCFs) associated with the calibration models, calibration lines are injected through one of the Pcals, the stages of the actuation, and, during O2, the control signal $d_{\rm ctrl}$.  The location of each injection is shown in Fig.~\ref{fig:DARM_loop}.  
The injections made in the actuation system, as well as one Pcal injection, are generally all placed together within a narrow ($\sim$2 Hz) frequency band, in order to justify an approximate direct comparison of the actuation injections to the Pcal injection.
This approximation, however, can still lead to significant systematic errors at times \cite{Sun:2020wke, VietsDissertation}.
Table~\ref{tab:callines} shows the purpose and approximate frequency of each calibration line.
\begin{table}[h!]
\centering
\caption{\label{tab:callines} Summary of the purpose of each calibration line. $^\ast$~denotes a specific parameter computation for a given line only applicable in O2.  $^\dag$~denotes a specific parameter computation for a given line only applicable in O3.}
\bigskip
\begin{tabular}{|| c | l | c ||}
\hline
\textbf{Line}  & \textbf{Purpose} & \textbf{Frequency} \\
\hline
\hline
$f_{\rm ctrl} \rule{0pt}{2.3ex} $ & Computation of $\kappa_{\rm PU}^\ast$ & 10 - 40 Hz \\

$f_{\rm T}$ \rule{0pt}{2.3ex} & Computation of $\kappa_{\rm T}$ & 10 - 40 Hz \\

$f_{\rm P}$ \rule{0pt}{2.3ex} & Computation of $\kappa_{\rm P}$ & 10 - 40 Hz \\

$f_{\rm U}$ \rule{0pt}{2.3ex} & Computation of $\kappa_{\rm U}$ & 10 - 40 Hz \\

$f^{\rm pc}_1$ \rule{0pt}{2.3ex} & Computation of $\kappa_{\rm T}$, $\kappa_{\rm PU}^\ast$ or $\kappa_{\rm P}^\dag$, $\kappa_{\rm U}^\dag$; $f_{\rm s}^\dag$ and $Q^\dag$ & 10 - 40 Hz \\

$f^{\rm pc}_2$ \rule{0pt}{2.3ex} & Computation of $\kappa_{\rm C}$  and $f_{\rm cc}$ & $\sim$400 Hz \\

$f^{\rm pc}_3$ \rule{0pt}{2.3ex} & Check on high-frequency calibration & $\sim$ 1 kHz \\

$f^{\rm pc}_4$ \rule{0pt}{2.3ex} & Computation of $f_{\rm s}^\ast$ and $Q^\ast$ & $\sim$ 8 Hz \\
\hline
\end{tabular}
\end{table}

In computing the TDCFs, the calibration of the Pcal is assumed to be accurate, that is, errors in Pcal laser power are uncompensated.  
In past cases of known beam clipping that reduced the power of the Pcal beam at the receiver module power sensor, the calibration configuration was modified to use the transmitter module power sensor instead.  
When it occurs, Pcal beam clipping causes an apparent reduction in the magnitude of the response function $\tilde{R}(f)$, resulting in inaccurate estimates of all the TDCFs.  
The calibration statevector, described in Appendix C of Ref.~\cite{hoft}, will flag times when the TDCFs stray outside of an expected range, which will alert calibrators to potential issues such as Pcal beam clipping that are impacting the calibration.

To compute the TDCFs, the amplitude and phase of each calibration line is measured in the error signal $d_{\rm err}$ and in the injection channels $x_{\rm pc}$, $x_{\rm T}$, $x_{\rm P}$, $x_{\rm U}$, and $x_{\rm ctrl}$ (O2 only) by demodulating each signal at the appropriate frequency.  
Ratios are then taken in order to compare the signal in $d_{\rm err}$ to the expected signal based on the injection channel and the calibration model at that frequency.  
The TDCF values are derived from these ratios, and full details of these calculations have been previously published in Refs.~\cite{Darkhan, hoft}.
After the TDCFs are computed, a running $\sim$2-minute median of each TDCF is calculated before being applied as described in Sec.~\ref{sec:filterupdates}.

During O2, the values of $f_{\rm s}$ and $Q$ at H1 estimated using the Pcal line at $f^{\rm pc}_4$ were subject to both large noisy fluctuations and systematic errors, evidenced by the fact that the average value of the quality factor $Q$ estimated using that line was negative.
However, we found that the calibration line at $f^{\rm pc}_1$ measures $f_{\rm s}$ and $Q$ with better precision and produces results that are generally consistent with reference-model measurements.
The improvement in precision is not surprising given the reduced seismic noise at the higher frequency, allowing for a calibration line with a higher signal-to-noise ratio.  
The apparent improvement in accuracy may be due to the increased impact of L2A2L cross-coupling at lower frequencies and the failure of the current sensing function model to correctly capture the frequency dependence of the low-frequency sensing function.  
For these reasons, calculations of $f_{\rm s}$ and $Q$ were based on the Pcal line at $f^{\rm pc}_1$ during O3.

However, compensation for time dependence in $f_{\rm s}$ and $Q$ in the reconstruction of $h(t)$ was omitted during O3 due to insufficient evidence of improvement in calibration accuracy by including time-dependence in $f_{\rm s}$ and $Q$.  
Again, this may have been due in part to the impact of L2A2L cross-coupling, but the approximations used to estimate the actuation TDCFs are also known to cause errors in the the estimates of $f_{\rm s}$ and $Q$.


\section{Compensating for TDCF Filters}
\label{sec:filterupdates}
Historically, the calibration of LIGO strain data has compensated for the TDCF multipliers, $\kappa_{\rm C}$, $\kappa_{\rm T}$, $\kappa_{\rm P}$, $\kappa_{\rm U}$, and $\kappa_{\rm PU}$ (O2 only)~\cite{Darkhan,hoft}.  
Here, we outline the methods developed in the calibration procedure to compensate for temporal variations in the interferometer response that require updating time-domain filters, known as TDCF filters, such as temporal variations in $f_{\rm cc}$, $f_{\rm s}$, and $Q$.  
The portion of the calibration procedure that applies the techniques described below uses finite impulse response (FIR) filters for all filtering processes, including the application of reference-model-based filters for $A$ and $C^{-1}$.

\subsection{Compensating for Temporal Variations in the Coupled Cavity Pole Frequency}
During O1 and O2, real-valued corrections for $\kappa_{\rm C}$, $\kappa_{\rm T}$, and $\kappa_{\rm PU}$ were applied to $h(t)$ by simply multiplying the components of $\Delta L_{\rm free}$ before summing:
\begin{eqnarray}
h(t) = \frac{1}{L} \Big( \kappa_{\rm T}(t) A_{\rm T} \ast d_{\rm ctrl}(t) &+& \kappa_{\rm PU}(t) A_{\rm PU} \ast d_{\rm ctrl}(t) \\
\nonumber
	&+& \frac{1}{\kappa_{\rm C}(t)} C^{-1} \ast d_{\rm err}(t) \Big).
\end{eqnarray}
Shortly after O2, an improved calibration was produced that additionally compensated for the time-dependence of the cavity pole frequency $f_{\rm cc}$ by applying and periodically updating a short correction filter just before the inverse sensing filter:
\begin{equation}
\label{eq:O2fccCorrection}
\Delta L_{\rm res}(t) = \frac{1}{\kappa_{\rm C}(t)} C^{-1}_{\rm static} \ast \left( \frac{1 + i f / f_{\rm cc}(t)}{1 + i f / f_{\rm cc}^{\rm static}} \right) \ast d_{\rm err}(t).
\end{equation}
where $f_{\rm cc}^{\rm static}$ refers to the cavity pole frequency of the static reference model for $C^{-1}$, and $f_{\rm cc}(t)$ is the cavity pole frequency as measured by the calibration pipeline using the methods discussed in Sec.~\ref{sec:kappas}.  
As can be inferred from Eq.~\eqref{eq:C}, the cavity pole frequency enters the inverse sensing function in the form $1 + i f / f_{\rm cc}(t)$.  
The term in parenthesis in Eq.~\eqref{eq:O2fccCorrection} is a short FIR filter used to compensate for time dependence in $f_{\rm cc}$; it both divides out the cavity pole component of $C^{-1}_{\rm static}$ and multiplies in the updated cavity pole component of the sensing function.
After measuring $f_{\rm cc}(t)$ and averaging its value over a specified time as described in Ref.~\cite{hoft}, correction FIR filters are created in regularly spaced intervals and smoothly transitioned into application for the $\Delta L_{\rm res}$ calculation by tapering out the previous correction filter and tapering in the current correction filter.

\subsection{Compensating for General TDCF Filters}
Since O2, this method has been further generalized to correct for the time dependence of $f_{\rm s}$ and $Q$ as well.  
For this purpose, we developed a filter-generation algorithm in the \texttt{gstlal-calibration} software package~\cite{gstlalcalibration} that takes as inputs an arbitrary number of zeros and poles, a gain, and a phase factor.  
The resulting filter is of the form
\begin{equation}
\tilde{\mathcal{F}}_{\rm corr}(f) = \frac{\prod_m \left(1 + i f / z_m\right)}{\prod_n \left(1 + i f / p_n\right)} K e^{2 \pi i f\tau},
\end{equation}
where the $z_m$ are the zero frequencies, the $p_n$ are the pole frequencies, $K$ is the gain of the filter, and $\tau$ is a time advance.  
The zeros and poles can be either read in as a time series or passed to the algorithm as constants, which is useful when dividing out zeros and poles from the static reference model.  
Additionally, the static reference-model filter or frequency-domain model can be multiplied by the correction filter in the frequency domain, so that the final product is the circular convolution $\mathcal{F} = \mathcal{F}_{\rm corr} \ast \mathcal{F}_{\rm static}$.  
This feature allows FIR filters for $C^{-1}$ and $A$ to be replaced, eliminating the need to have two filters in series and allowing for reduced filter latency and higher quality application of frequency-dependent corrections.  
The procedure used to update the inverse sensing filter is as follows:
\begin{enumerate}
\item In the frequency domain, compute the correction filter
\begin{eqnarray}
\tilde{C}^{-1}_{\rm corr}(f) &=& \frac{1}{\kappa_{\rm C}(t)} \frac{1 + i f / f_{\rm cc}(t)}{1 + i f / f^{\rm static}_{\rm cc}} \\
\nonumber
	&\times& \frac{f^2 + f_{\rm s}^2(t) - i f f_{\rm s}(t) / Q(t)}{f^2 + (f^{\rm static}_{\rm s})^2 - i f f^{\rm static}_{\rm s} / Q^{\rm static}}.
\end{eqnarray}
In the frequency domain, the number of samples required to produce the correction filter is one more than half the length of the static time-domain filter $C^{-1}_{\rm static}$.  
The transformation into the time domain via an inverse discrete Fourier transform (step 5) then produces a filter of the desired length.\footnote{The method outlined here is general and includes the possibility of applying a correction for the the time-dependent $f_s$ and $Q$ parameters.  However, in practice neither of these time-dependent parameters has been applied in released calibrated data at the time of writing.}
\item In the frequency domain, multiply the correction filter by the frequency-domain static model.  If this is not provided to the algorithm, it is computed from the static filter using a discrete Fourier transform.\footnote{The static filter contains an added delay which must be removed before the multiplication.}
\item Add a delay to the filter of half the length of the filter to ensure that the resulting time-domain filter is centered in time.  Assuming the filter has an even length, this can be done simply by negating every other value in the frequency-domain filter, starting after the DC component and ending before the Nyquist component. This is equivalent to multiplying each frequency-domain value by $e^{-\pi i f \tau_{\rm filt}}$, where $\tau_{\rm filt}$ is the temporal duration of the time-domain filter.
\item Take the inverse real discrete Fourier transform of the frequency-domain filter to produce a time-domain filter equal in length to the static filter $C^{-1}_{\rm static}$.  The algorithm used assumes that the output is to be real, and that the input is only the first half of a conjugate-symmetric array, containing the frequency-domain filter only from the DC component to the Nyquist component.
\item Apply a window function to the time-domain FIR filter so that it falls off smoothly at the edges.  A Tukey window was used for this purpose during O3.  In the future, a Slepian window will be used instead to maximize energy concentration in the main lobe of the filter's frequency response, resulting in a significant improvement in filtering quality.
\item Pass the updated filter to an algorithm which applies FIR filters and smoothly handles filter updates by using half of a Hann window to taper out the old filter and taper in the new filter \cite{Leo}.  During a transition of duration $t_{\rm trans}$ beginning at $t_0$, the output is therefore
\begin{eqnarray}
\Delta L_{\rm res}(t) &=& \cos^2 \left( \frac{\pi}{2} \cdot \frac{t - t_0}{t_{\rm trans}} \right) C^{-1}_{\rm old} \ast d_{\rm err}(t) \label{eq:tdwhiten} \\
\nonumber
 &+& \sin^2 \left( \frac{\pi}{2} \cdot \frac{t - t_0}{t_{\rm trans}} \right) C^{-1}_{\rm new} \ast d_{\rm err}(t).
\end{eqnarray}
During O3, the transition time $t_{\rm trans}$ was 2 seconds for the inverse sensing filter.
\end{enumerate}

A similar procedure can also been used to apply corrections that include both magnitude and phase to the actuation filters $A_{\rm T}$, $A_{\rm P}$, and $A_{\rm U}$.  It is possible to compensate for temporal variations in the variable time advances $\tau_{\rm T}$, $\tau_{\rm P}$, and $\tau_{\rm U}$, which have shown occasional changes, using a linear-phase FIR filter.  The time-varying actuation filters are therefore
\begin{equation}
\tilde{A}_i(f) = \kappa_i e^{-2 \pi i f \tau_i} \tilde{A}^{\rm static}_i,
\end{equation}
where $i \in$ \{T, P, U\}.

\subsection{Two-Tap Filters for Time-Varying Zeros}
The TDCFs of the inverse sensing function consist of only a gain, $\kappa_{\rm C}$, and three zeros: $f_{\rm cc}$ and the two zeros associated with $f_{\rm s}$ and $Q$.
We developed an alternative method to compensate for time-dependent gains and zeros, since it is possible to model a zero (and a gain) using an FIR filter with only two taps.
For a gain $K$ and a single zero at frequency $f_{\rm z}$, the filter coefficients $a_i$ are
\begin{equation}
\begin{array}{ll}
a_0 &= K \left(\frac{1}{2} + \frac{f_{\rm samp}}{2 \pi f_{\rm z}}\right) \\
a_1 &= K \left(\frac{1}{2} - \frac{f_{\rm samp}}{2 \pi f_{\rm z}}\right) ,
\end{array}
\end{equation}
where $f_{\rm samp}$ is the sampling frequency.
A time-varying filter modeling the impact of $\kappa_{\rm C}$ and $f_{\rm cc}$ can be applied in series with a static filter associated with the time-independent portion of the inverse sensing function.

The primary advantage of using this method is a significant reduction in computational cost.
The adaptive filtering used to implement Eq.~\eqref{eq:tdwhiten} is the most computationally expensive single-threaded process in the calibration procedure, consuming about half of the computational power of a CPU, if using the hardware available during O3 to apply an adaptive filter of one-second duration at a sample rate of 16 kHz.
This contributed $\sim$0.5 s of latency, that is, half of the duration of the one-second buffers of input data used during low-latency operation.
Replacing this with a static filter and a very short adaptive filter reduces computational time by almost 50\%, reducing latency by $\sim$0.25 s.  For the same reason, this would also reduce the total computation time required to produce high-latency calibrated data.
It is therefore likely that this method will be implemented during Advanced LIGO's fourth observing run.

\begin{figure}[h!]
\begin{center}
\includegraphics[width=\columnwidth]{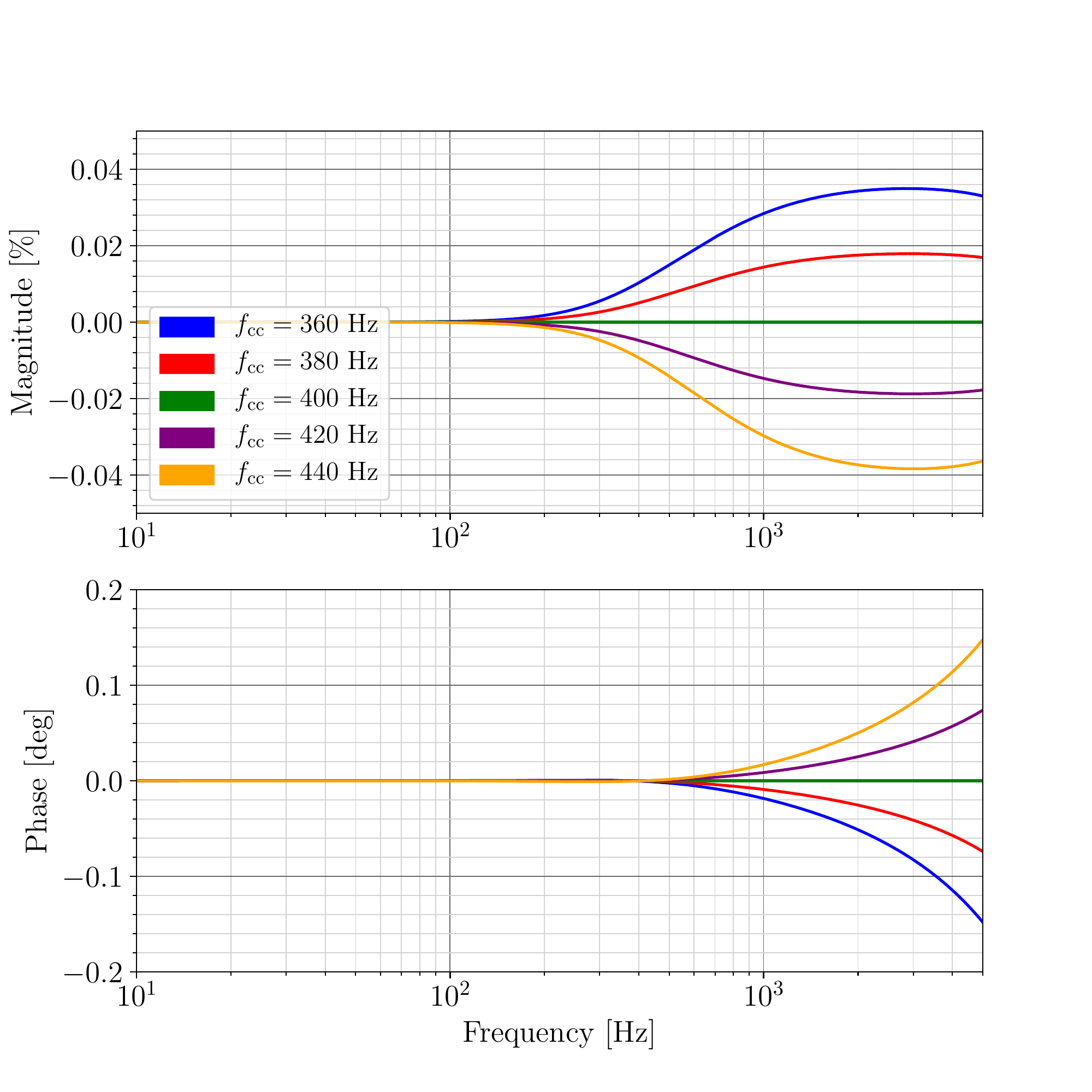}
\end{center}
\caption{Errors induced in the inverse sensing function by using a two-tap filter to compensate for temporal variations in $f_{\rm cc}$, assuming a static error compensation with a value of $f_{\rm cc} = 400$~Hz has already been applied to the inverse sensing function.  The errors shown for different values of $f_{\rm cc}$ therefore show the remaining error caused by the fact that $f_{\rm cc}$ does not remain statically at a value of 400~Hz.  Fluctuations in $f_{\rm cc}$ larger than 20~Hz are uncommon, so these results show dramatic maximal errors that would be induced in this situation, which remain below 0.04\% in magnitude and 0.2 degrees in phase.
\label{fig:twoTap}}
\end{figure}

There is a small error introduced in magnitude and phase when using the two-tap filter compared to the full adaptive filter.
The error is mostly due to an uncompensated time-delay equal to half of a sample period.
This can be corrected by compensating for the error in the static inverse sensing filter.  
To compute the correction, we compute a static inverse sensing filter based on the full static inverse sensing function model, and then divide out the exact frequency response of a two-tap filter corresponding to the static reference model value of $f_{\rm cc}$.  
The static filter exactly compensates for the error of the two-tap filter if $f_{\rm cc}$ is equal to its nominal value.  
Although this error changes slightly as $f_{\rm cc}$ changes, the magnitude of this change is so small that compensating for the error as though it were static results in negligible errors in the inverse sensing function, as shown in Fig.~\ref{fig:twoTap}.  

Compensation for additional zeros using this method can be achieved simply by convolving additional two-tap filters to produce a longer filter.  
Therefore, compensation for the gain and three time-dependent zeros of the inverse sensing function requires a four-tap filter.

\section{Impact on Calibration Accuracy}
\label{sec:calibrationAccuracy}

\begin{figure*}[t]
\begin{center}
\includegraphics[width=\textwidth]{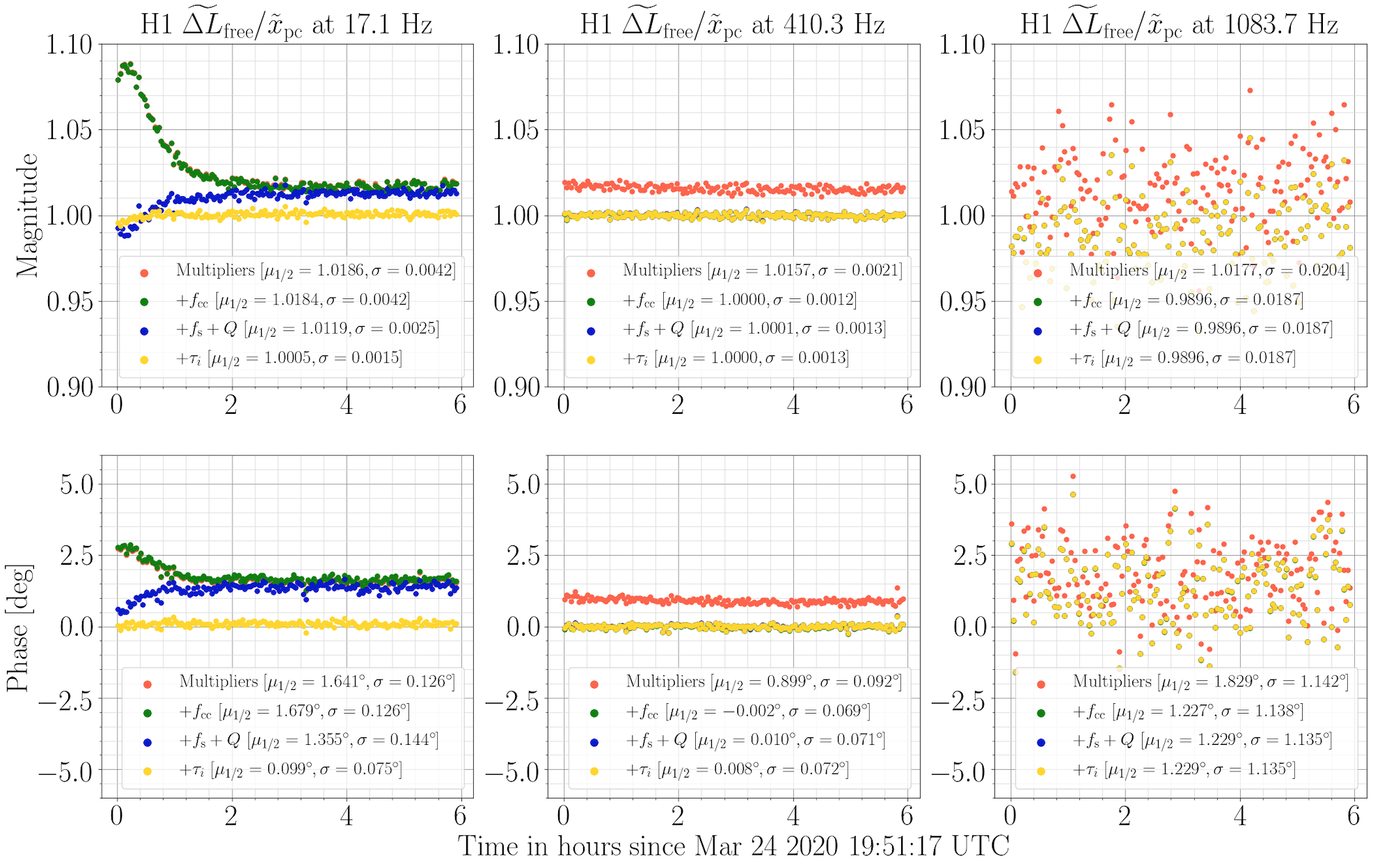}
\end{center}
\caption{The ratio $\Delta L_{\rm free}(f) / x_{\rm pc}(f)$ at three Pcal line frequencies for three versions of calibrated data for H1.  150 seconds of $\Delta L_{\rm free}$ data and $x_{\rm pc}$ data were demodulated before taking the ratio to produce each point.  The red points (labeled ``Multipliers") represent calibrated data that was corrected for the time dependence of $\kappa_{\rm T}$, $\kappa_{\rm P}$, $\kappa_{\rm U}$, and $\kappa_{\rm C}$, requiring no filter updates.  The green points (labeled ``$+ f_{\rm cc}$") show improved accuracy resulting from additionally compensating for time-dependence in $f_{\rm cc}$.  The blue points (labeled ``$+ f_{\rm s} + Q$") indicate compensation for time dependence in $f_{\rm s}$ and $Q$.  The yellow points (labeled ``$+ \tau_i$") indicate compensation for all known time-dependence.  Some colors appear below other colors in certain figure panels due to the alignment of different results.
\label{fig:pcal2darm}}
\end{figure*}

\begin{figure}[!h]
    \centering
    \includegraphics[width=\columnwidth]{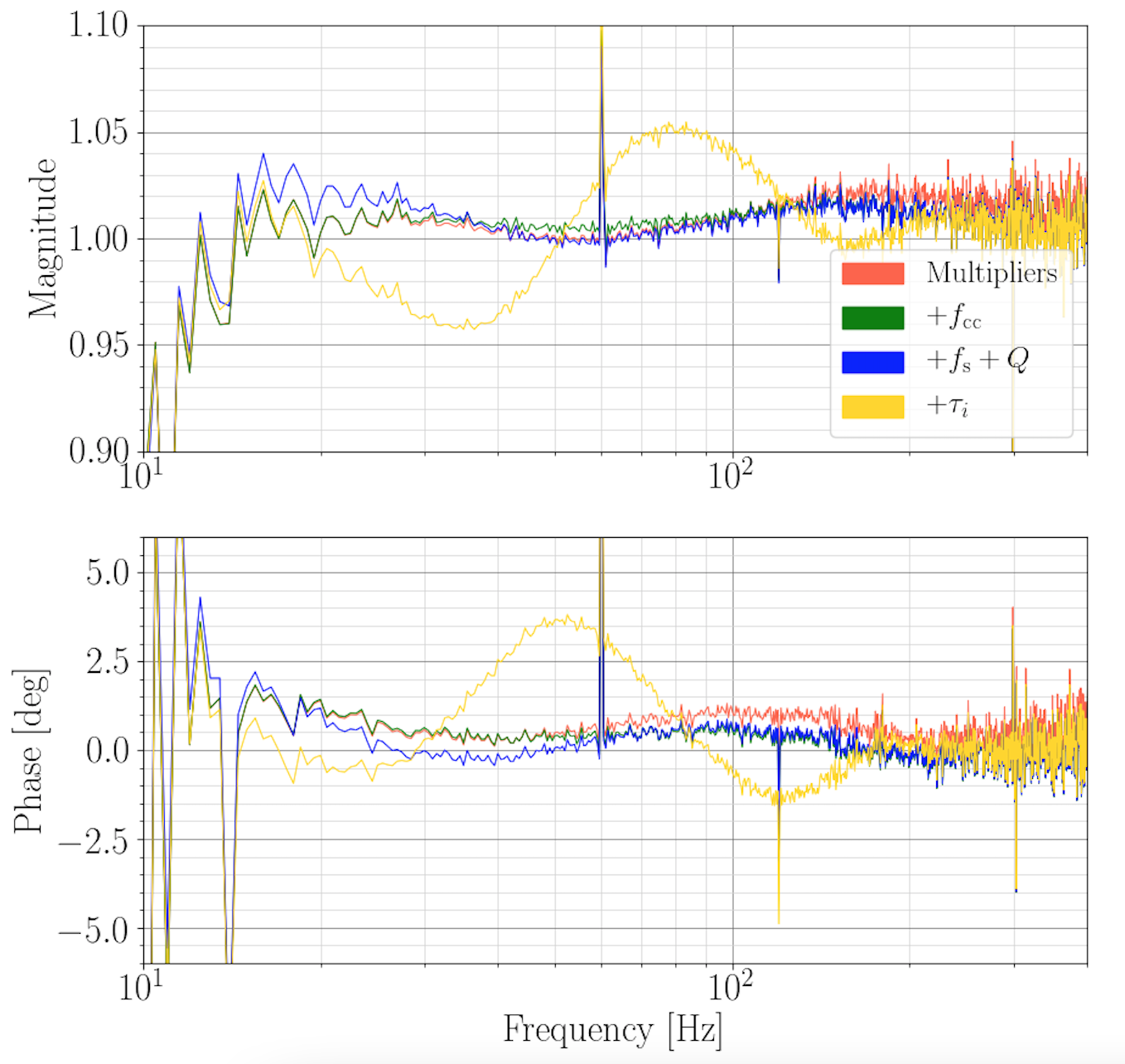}
    \caption{The transfer function $\Delta L_{\rm free}(f) / x_{\rm pc}(f)$ computed for four versions of calibrated data for H1.  Each transfer function was produced from 178 seconds of data during a Pcal broadband injection on 2020-03-23.  This shows an instance in which compensation for the $\tau_i$, although it improves accuracy at the Pcal lines, results in increased systematic errors at most frequencies.
\label{fig:pcalBroadbandManual}}
\end{figure}

\begin{figure}[!h]
    \centering
    \includegraphics[width=\columnwidth]{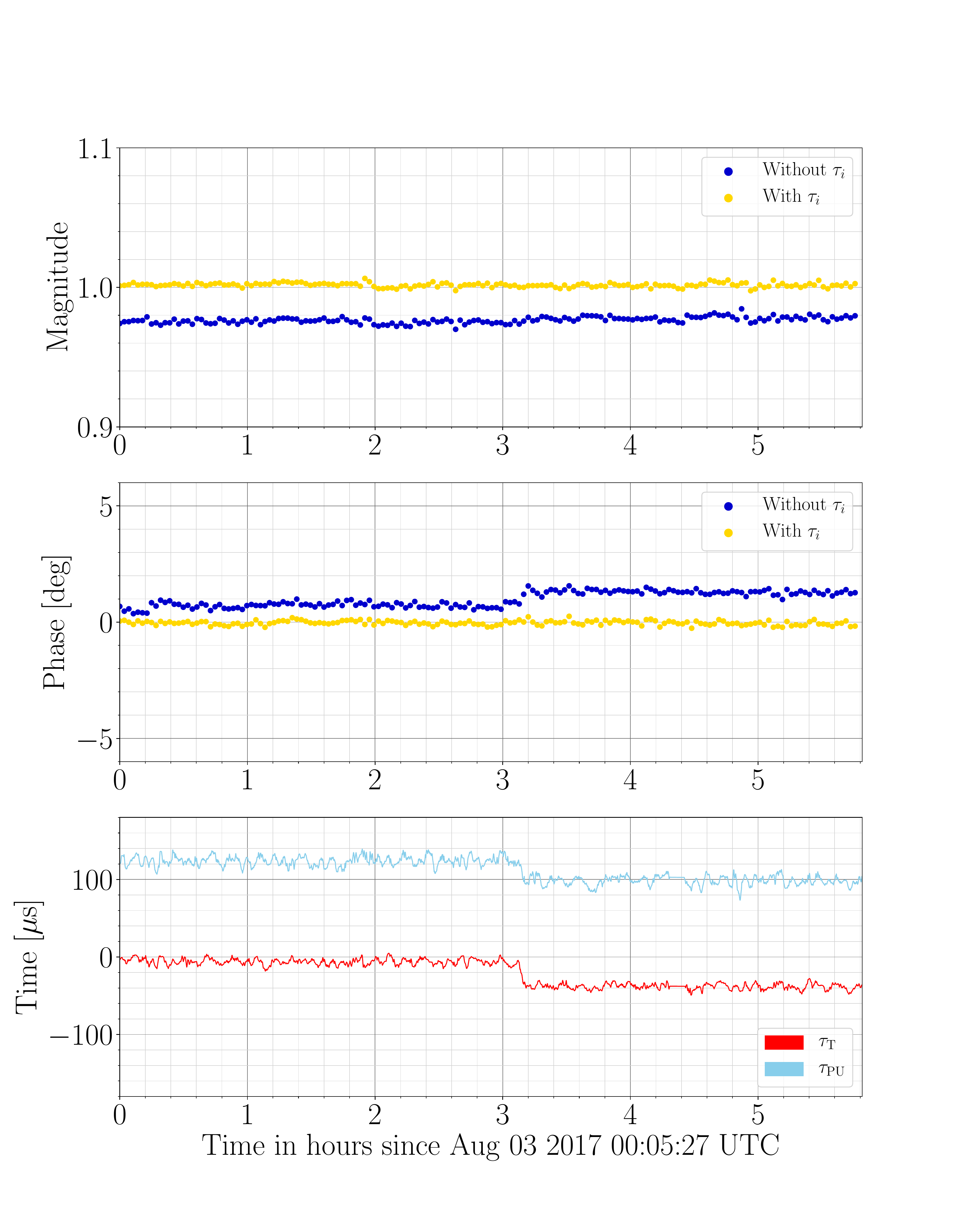}
    \caption{A time series of the $\tau_i$ (bottom plot) and ratio $\Delta L_{\rm free}(f^{\rm pc}_1) / x_{\rm pc}(f^{\rm pc}_1)$ (top two plots) at the Pcal line frequency $f^{\rm pc}_1$ = 36.7 Hz.  The sudden simultaneous shift in the $\tau_i$ and the phase of $\Delta L_{\rm free}(f^{\rm pc}_1) / x_{\rm pc}(f^{\rm pc}_1)$ indicates a change to a computational time delay, whose negative impact on calibration accuracy is corrected by compensating for time dependence in the $\tau_i$.
\label{fig:actTiming}}
\end{figure}

To assess the impact of compensating for frequency-dependent temporal variations, we compared calibrated $h(t)$ data to the calibration Pcal signal at the Pcal line frequencies.  
Since the calibration lines run continuously during observation, this comparison can be tracked for long periods of time to test the accuracy and stability of the calibration at the Pcal line frequencies.  
The impact of TDCF filters in $C$ and $A$ was significantly larger at H1 than at L1 during O3.  Therefore, the data used for this analysis was taken from H1.

Fig.~\ref{fig:pcal2darm} shows a time series of the magnitude and phase of the ratio $\Delta L_{\rm free}(f) / x_{\rm pc}(f)$ at three Pcal lines.  
As expected, correcting for all known time dependence yields the most accurate result at the Pcal lines, with systematic errors generally less than 1\% in magnitude and 1$^{\circ}$ in phase.  
This result, however, only indicates that the solution and application of the TDCFs correctly enforces agreement at the calibration lines used to compute the TDCFs and does not necessarily imply correctness in the time-dependent calibration model.

The time-dependence of the coupled cavity pole $f_{\rm cc}$ is the most significant source of systematic error at the $f^{\rm pc}_2=410.3$~Hz Pcal line, where a systematic error of 1\% in magnitude and 1$^{\circ}$ in phase is reduced to a negligible level by compensating for the time dependence of $f_{\rm cc}$.
The time dependence of $\tau_{\rm T}$, $\tau_{\rm P}$, $\tau_{\rm U}$, $f_{\rm s}$, and $Q$ is primarily impactful at lower frequencies.  
The systematic error at $f^{\rm pc}_1 = 17.1$~Hz seen in Fig.~\ref{fig:pcal2darm} is reduced to negligible levels by compensating for time-dependence in these parameters.

In order to assess the impact of compensating for frequency-dependent temporal variations on calibration accuracy across a broad range of frequencies, we used broadband Pcal injections to compute the transfer function $\Delta L_{\rm free}(f) / x_{\rm pc}(f)$ from 10 Hz to 400 Hz.
Fig.~\ref{fig:pcalBroadbandManual} shows broadband injection results from March 23, 2020.

Due to the fact that driving DARM motion becomes more difficult with increasing frequency, the transfer function estimate becomes noisier at higher frequencies.  
At the higher frequencies, a small improvement in systematic error is seen, as expected, due to compensation for time dependence in $f_{\rm cc}$.  Since $f_{\rm cc}$ was fairly close to its nominal value at the time of the injection ($f_{\rm cc}^{\rm static} = 411$~Hz, $f_{\rm cc}(t) = 420$~Hz), the improvement is small.  
The impact is expected to be larger at times when $f_{\rm cc}$ undergoes more noticeable change, such as in the hours following the interferometer acquiring lock, which encompasses the time during which the interferometer optics are thermalizing.  

A more obvious impact is the increase in systematic error caused by compensating for time dependence in the $\tau_i$.  
This seems to indicate that the deviation of $\tau_i$ from zero was dominated by systematic errors in their estimates at this time, as was the case throughout much of O3.  
For this reason, for most of O3, compensation for time dependence in the $\tau_i$ was omitted.  
Further discussion of why estimates of the $\tau_i$ are known to deviate from zero can be found in Appendix~\ref{app:tau}.
In the future, this problem can be mitigated through use of the exact solution for the TDCFs developed in Ref.~\cite{VietsDissertation}.

As opposed to what was observed in O3, evidence from O2 data generally indicates an improvement in calibration accuracy due to compensation for time dependence in the $\tau_i$ parameters, as suggested by Fig.~\ref{fig:actTiming}.  
This figure shows a time series of the $\tau_i$ and the ratio $\Delta L_{\rm free}(f^{\rm pc}_1) / x_{\rm pc}(f^{\rm pc}_1)$, including a moment at which a computational time-delay in the DARM loop suddenly increased by $\sim$\SI{30}{\micro \second}.
The data that is compensated for time dependence in the $\tau_i$ is immune to this sudden shift.
By employing a calculation for $\tau_i$ that is free of many of the systematic errors that plagued this parameter in O3~\cite{VietsDissertation}, we would be able to ensure improved calibration accuracy when compensating for temporal variations in $\tau_i$.

\section{Impact on Astrophysical Analyses}
\label{sec:astrophysics}

\begin{figure*}
\centering
\begin{subfigure}{.5\textwidth}
  \centering
  \includegraphics[width=\linewidth]{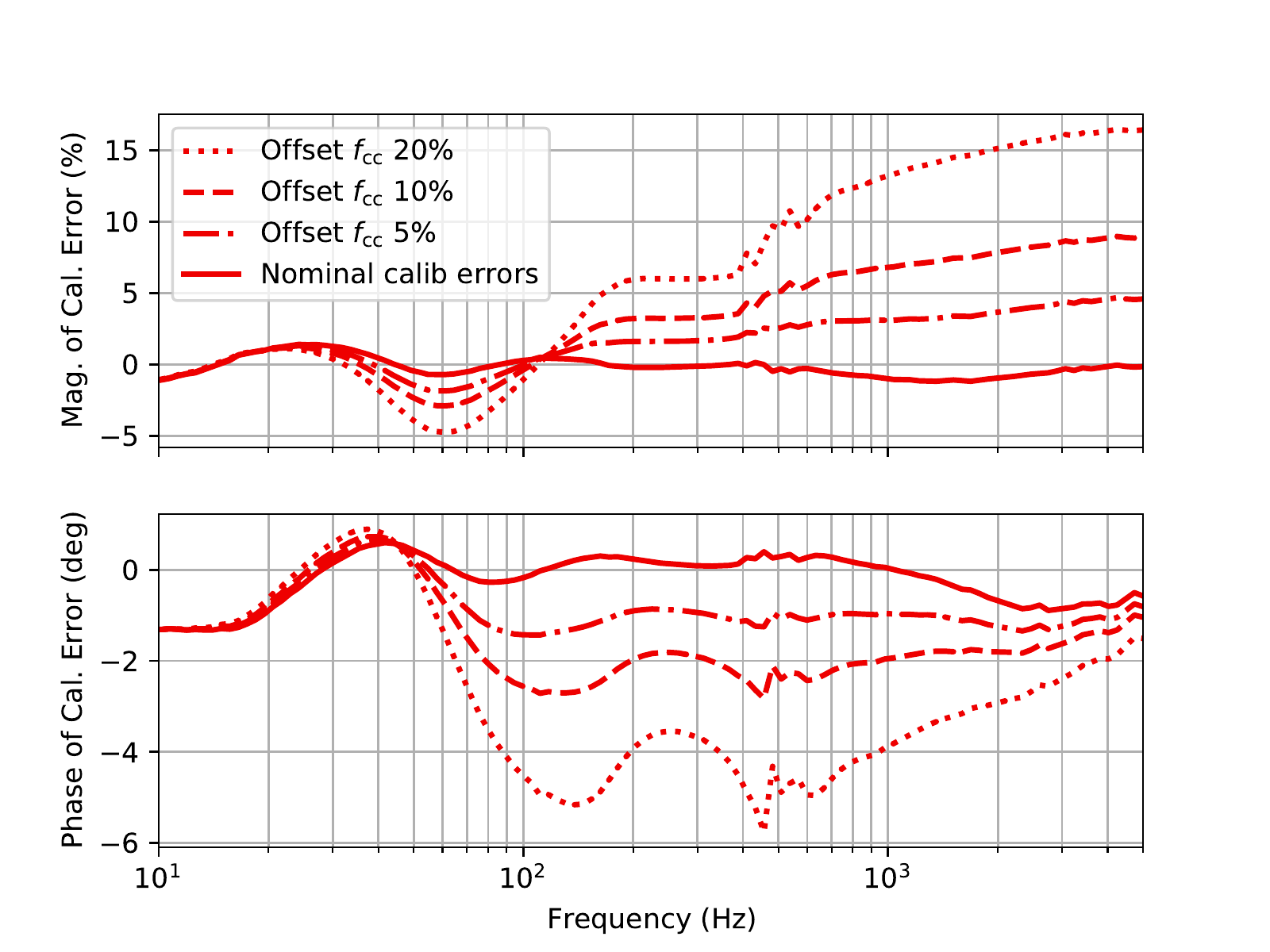}
  \label{fig:sub1}
\end{subfigure}%
\begin{subfigure}{.5\textwidth}
  \centering
  \includegraphics[width=\linewidth]{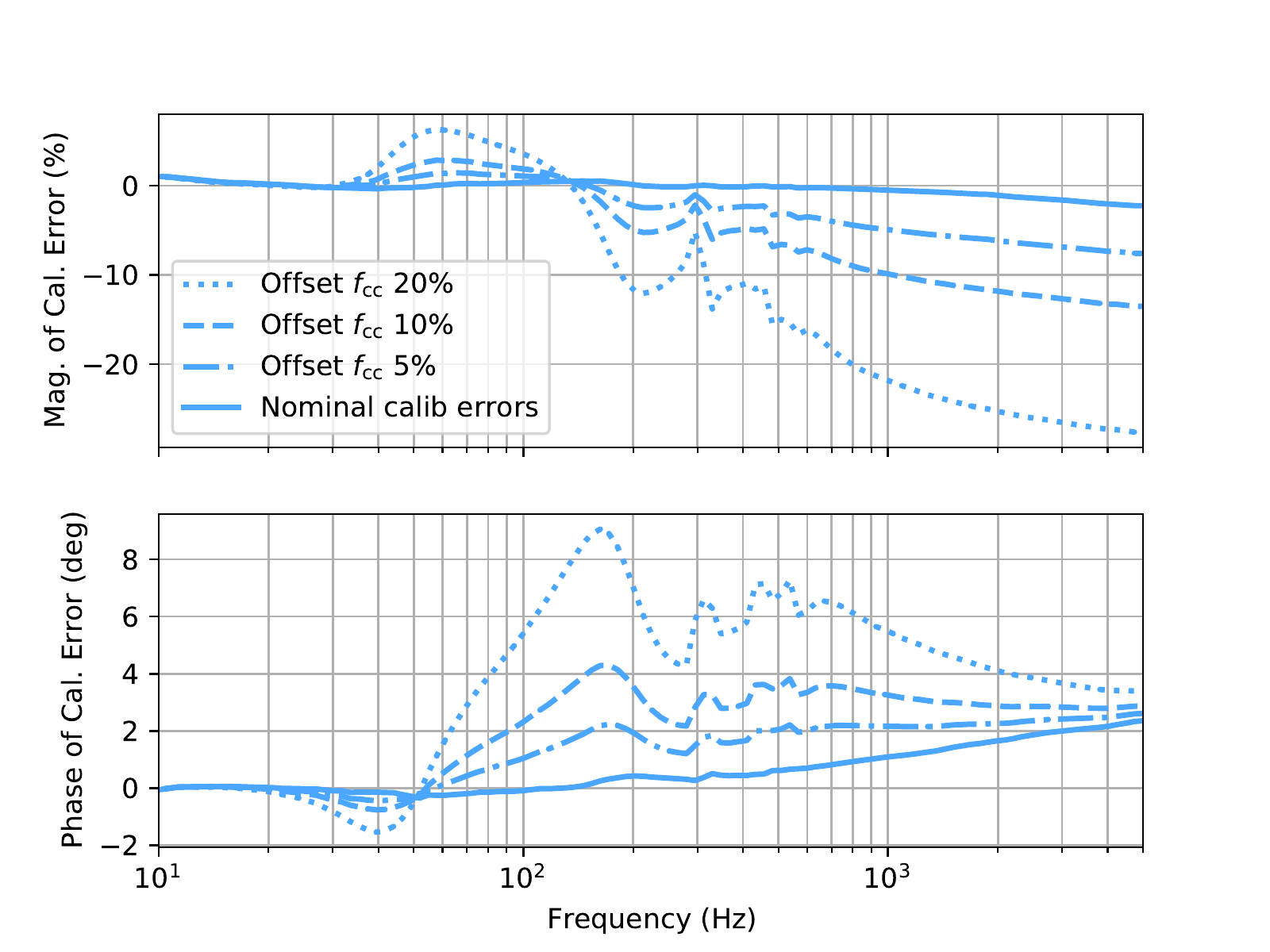}
  \label{fig:sub2}
\end{subfigure}
\caption{\label{fig:fc_miscal} Magnitude (top plots) and phase (bottom plots) of calibration errors at H1 (left) and L1 (right) when purposely offsetting the cavity pole frequency $f_{\rm cc}$ by $5\%$, $10\%$, and $20\%$ from its nominal value of 410.6 Hz (H1) or 454.0 Hz (L1).  For the H1 results, $f_{\rm cc}$ was increased by $5\%$, $10\%$, and $20\%$, and for the L1 results, $f_{\rm cc}$ was decreased by these percentages.  The solid line in each figure indicates the calibration errors present for nominal choices of all TDCFs.}
\end{figure*}

\begin{figure*}
\centering
\begin{subfigure}{.5\textwidth}
  \centering
  \includegraphics[width=\linewidth]{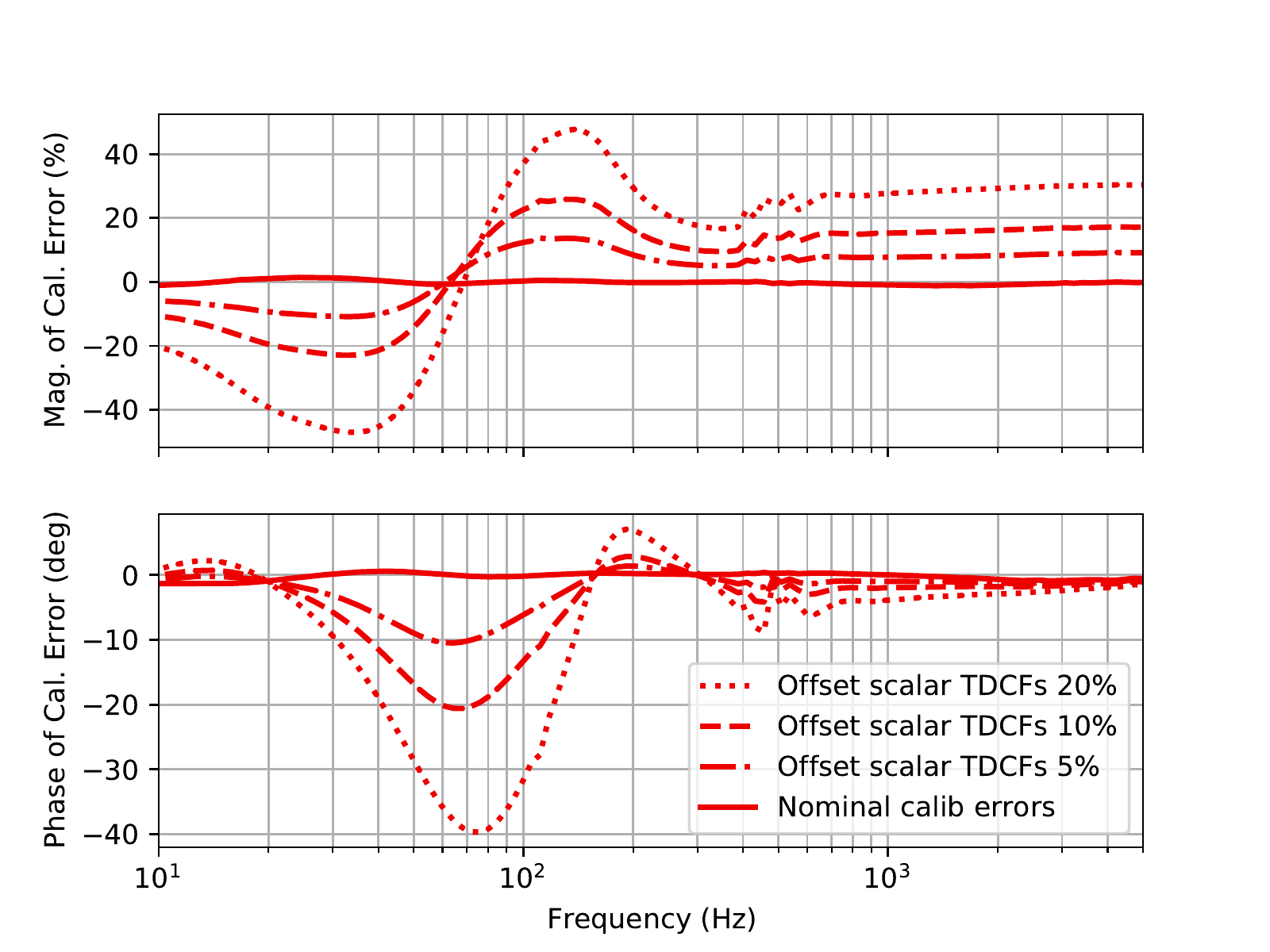}
  \label{fig:sub1}
\end{subfigure}%
\begin{subfigure}{.5\textwidth}
  \centering
  \includegraphics[width=\linewidth]{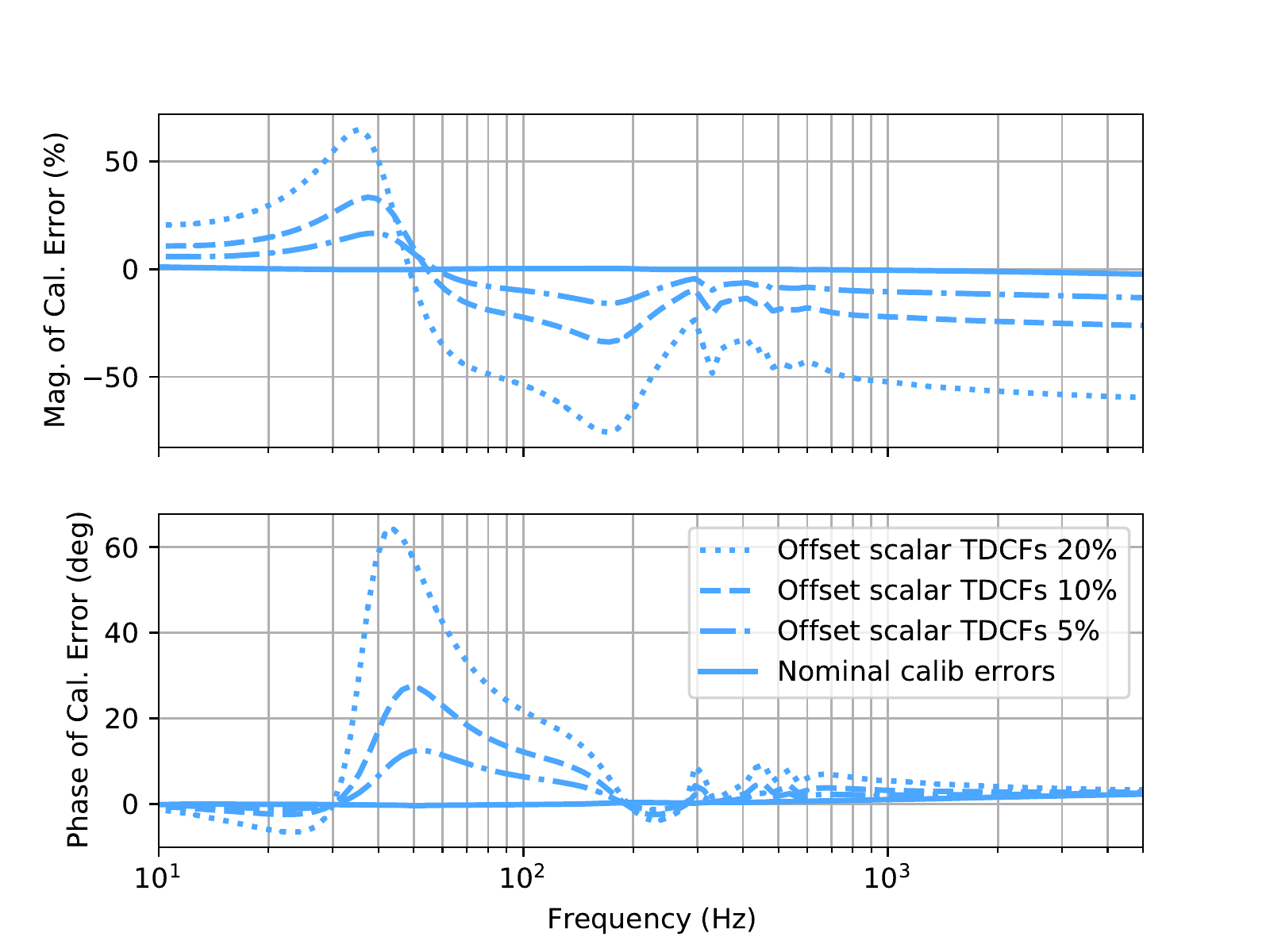}
  \label{fig:sub2}
\end{subfigure}
\caption{\label{fig:all_miscal} Magnitude (top plots) and phase (bottom plots) of calibration errors at H1 (left) and L1 (right) when purposely offsetting the TDCF multipliers ($\kappa_{\rm c}$, $\kappa_{\rm T}$, $\kappa_{\rm P}$, and $\kappa_{\rm U}$) by $5\%$, $10\%$, and $20\%$.  For the H1 results, the TDCF multipliers were increased by $5\%$, $10\%$, and $20\%$, and for the L1 results, the TDCF multipliers were decreased by these percentages.  The solid line in each figure indicates the calibration errors present for nominal choices for all TDCFs.  The total systematic error induced by adjusting all of the TDCF multipliers can be quite large, relative to the individual adjustment of each TDCF multiplier.  For example, a 20\% decrease in all TDCF multipliers leads to errors as large as 80\% in magnitude and $>60$ degrees in phase in certain frequency ranges.}
\end{figure*}

Systematic errors in Advanced LIGO’s calibrated data, such as those introduced by not compensating for the time-dependence of some calibration model parameters, have the potential to impact astrophysical results that flow from the reconstructed strain data.  
Calibration errors have a complex frequency structure, especially when frequency-dependent temporal variations are ignored.  
Fig.~\ref{fig:fc_miscal} shows examples of calibration errors with different intentional offsets of the time-dependent cavity pole frequency parameter $f_{\rm cc}$.  
Fig.~\ref{fig:all_miscal} shows examples of calibration errors with different intentional offsets of TDCF multipliers.  
With the exception of the nominal calibration error in each plot, these figures represent simulations of potential calibration errors rather than realized calibration errors from previous observing runs. 
A full discussion of the realized calibration errors in previous observing runs, including both the calibration systematic error and its associated uncertainty, can be found in previous publications~\cite{Craig,Sun:2020wke,Sun:2021qcg}.

Since systematic errors in LIGO’s calibrated data impact the reconstructed strain and astrophysical source parameters impact the physical strain, potentially in similar ways, we sought to address the question of how much the systematic error caused by not compensating for changes in the TDCF filters could impact astrophysical results.
We investigated a specific scenario related to this question by studying the impact of the additional systematic error caused by not compensating for temporal changes in the coupled cavity pole frequency $f_{\rm cc}$ on the source parameter estimation of a binary neutron star (BNS) system.
In addition, we separately studied how the systematic error caused by not compensating for TDCF mulitipliers would impact the source parameter estimation of the same BNS system.
Previous studies have developed sophisticated frameworks to fully incorporate the calibration systematic error and its associated uncertainty into the parameter estimation procedure.
We have used a much simpler framework than those developed in these previous studies~\cite{Payne:2020myg, Vitale:2020gvb}.  
We apply systematic calibration errors, both including and not including compensation for TDCFs, to a simulated signal in order to study biases in the parameter estimation results that are a consequence of the systematic error only.

In general, during O2 and O3, the systematic error present in the calibrated strain data is estimated numerically by producing many iterations of possible response functions.  
The set of possible response functions are produced by using a combination of a Markov Chain Monte Carlo (MCMC) method and Gaussian process regression (GPR).  
The MCMC is used to estimate the DARM model parameters. 
The maximum likelihood values obtained from the MCMC are then used to construct a nominal DARM model.  
With the nominal DARM model in hand, a GPR is used to estimate any remaining deviations of the DARM model from the full interferometer response using measurements obtained with the Pcal. 
Possible response functions are generated by sampling from the distribution of the MCMC and GPR results while also accounting for the TDCF uncertainty at a given time and the Pcal uncertainty.  
For more detail on estimating Advanced LIGO calibration errors and uncertainties in O2 and O3 see Refs.~\cite{Craig,Sun:2020wke,Sun:2021qcg}.

To study the potential impact of systematic calibration errors on the estimation of astrophysical parameters from a binary coalescence event~\cite{PhysRevD.49.2658}, we developed a range of calibration systematic error estimates that included different deviations of TDCFs from their nominal values.  
The nominal calibration error estimate that each manipulation was based on was chosen from June 11, 2019 (GPS time 1244307456) for H1 and March 27, 2019 (GPS time 1237745764) for L1.  
We produced six modifications to this nominal calibration error estimate for each detector.  
One set of modifications focused on manipulating only the TDCF for the coupled cavity pole frequency $f_{\rm cc}$, and the other set of modifications manipulated all of the TDCF multipliers ($\kappa_{\rm c}$, $\kappa_{\rm T}$, $\kappa_{\rm P}$, and~$\kappa_{\rm U}$).  

Fig.~\ref{fig:fc_miscal} shows the nominal systematic calibration errors in H1 and L1 as well as the calibration errors resulting from three manipulations of $f_{\rm cc}$.  
The cavity pole frequency parameter was intentionally offset by $5\%$, $10\%$ and $20\%$ from its nominal value and then the calibration systematic error was computed for the resulting response functions.  
The cavity pole frequency was increased by these percentages in H1 and decreased by these percentages in L1 in order to mimic a maximal relative calibration error, manifesting in a similar way to a relative timing error, between each set of detector strain data.  
Since the relative timing of the signal between H1 and L1 is the primary contributor to the sky localization of a detected signal, we wanted to study how offsetting these parameters in opposite ways at H1 and L1 would impact the sky localization.
In Ref.~\cite{maddiethesis}, which this work builds on, results are shown for an offset of the TDCFs in the same direction for H1 and L1, which found calibration errors to have a minimal effect on the measured source parameter distributions.
Here we only highlight the more interesting results with offsets in opposite directions at H1 and L1.

Fig.~\ref{fig:all_miscal} similarly shows the nominal systematic calibration error in H1 and L1 as well as the calibration errors resulting from three manipulations of the TDCF multipliers.
Similarly to above, the manipulations involved intentionally offsetting each TDCF multiplier from its nominal value by $5\%$, $10\%$, and $20\%$ in opposite directions for H1 and L1.  

For comparison to expected physical deviations, it is rare for the coupled cavity pole to deviate from its nominal value by more than $5\%$.  
The TDCF multipliers, however, are known to deviate by as much as $\sim10\%$ from their nominal values.
The set of simulated calibration errors used in this study should therefore encompass both realistic and extreme calibration error situations.

In order to test the impact of the above calibration systematic errors on the estimation of source parameters of a binary system, we applied the calibration systematic errors to a simulated BNS gravitational-wave signal as well as the simulated noise floor (power spectral density) before performing parameter estimation using the \texttt{lalinference} software package.
We specifically used \texttt{lalinference}'s MCMC sampler for all simulations presented here, the details of which can be found in Ref.~\cite{PhysRevD.91.042003}. 
The MCMC sampler produces posterior probability distributions (or just posteriors for short) for the signal's source parameters.
Calibration errors were not marginalized over to produce the posteriors.

We chose to add no synthetic noise to the simulated BNS signal (sometimes referred to as ``injecting into zero-noise" \cite{Rodriguez_2014}) in order to isolate biases in the source parameter estimation caused by the calibration systematic error from the varying effects of a given instance of synthetic noise \cite{Nissanke_2010}. 
Since the likelihood calculation involves a division by the power spectral density (PSD) of the synthetic noise, this approach does still incorporate the general properties of the noise in the analysis.
We used the simulated Advanced LIGO sensitivity PSD within the likelihood calculation \cite{LIGOpsd}.
The waveform approximant used both for the simulated signal and for the template was \texttt{TaylorF2} carried out to 3.5 post-Newtonian order \cite{PhysRevD.80.084043,PhysRevLett.112.101101,Wade:2014vqa}.  
The starting frequency used for this analysis was 40~Hz.
The parameters of the simulated BNS source were chosen from a random distribution to have masses $m_1=1.523194 \ M_\odot$ and $m_2 = 1.522147 \ M_\odot$, zero spin, and tidal deformability parameters $\Lambda_1 = 311.368$ and $\Lambda_2 = 312.666$ \cite{PhysRevD.77.021502,PhysRevD.81.123016}.  
We performed parameter estimation on this simulated signal placed at two different distances, resulting in two different signal-to-noise ratios (SNRs).  
The high-SNR signal had a simulated distance of 58~Mpc, which resulted in a network SNR of 56.  
The low-SNR signal had a simulated distance of 200~Mpc, which resulted in a network SNR of 16.
Our template waveform included 11 parameters: 4 intrinsic parameters (two mass parameters and two tidal deformaibility parameters) and 7 extrinisic parameters.
We do not consider spin in our template waveforms.
We used uniform priors in component tidal deformability between $0<\Lambda_{1,2}<3000$ and a uniform prior distribution in volume to a distance of 300~Mpc$^3$.
The rest of our priors were consistent with those outlined in Sec.~IIIB of Ref.~\cite{Wade:2014vqa}.
For this study we used a two-detector network consisting of H1 and L1, since we were focused on calibration errors specific to the LIGO detectors.

\begin{figure*}
\centering
\begin{subfigure}{.5\textwidth}
  \centering
  \includegraphics[width=\linewidth]{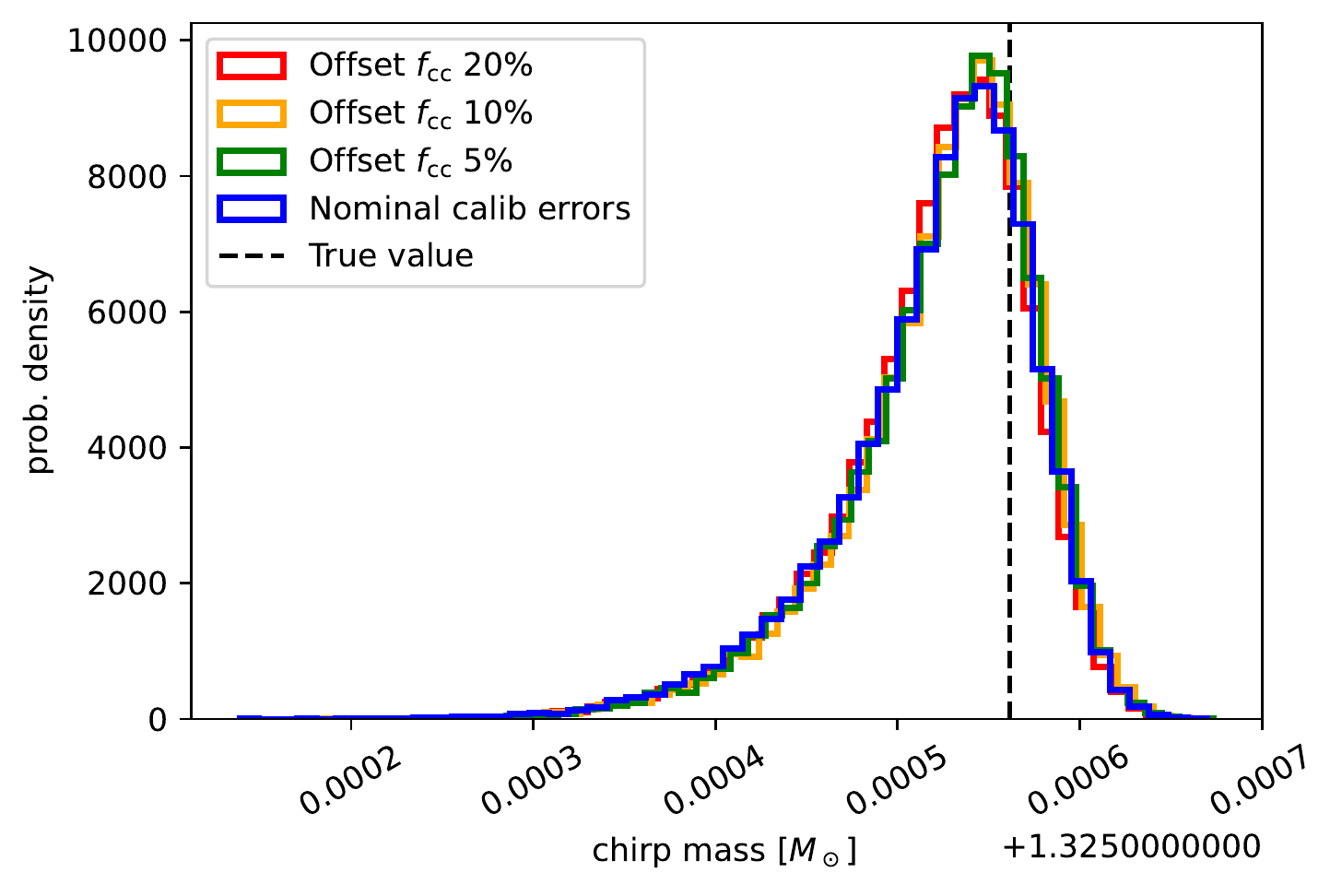}
  \label{fig:sub1}
\end{subfigure}%
\begin{subfigure}{.5\textwidth}
  \centering
  \includegraphics[width=\linewidth]{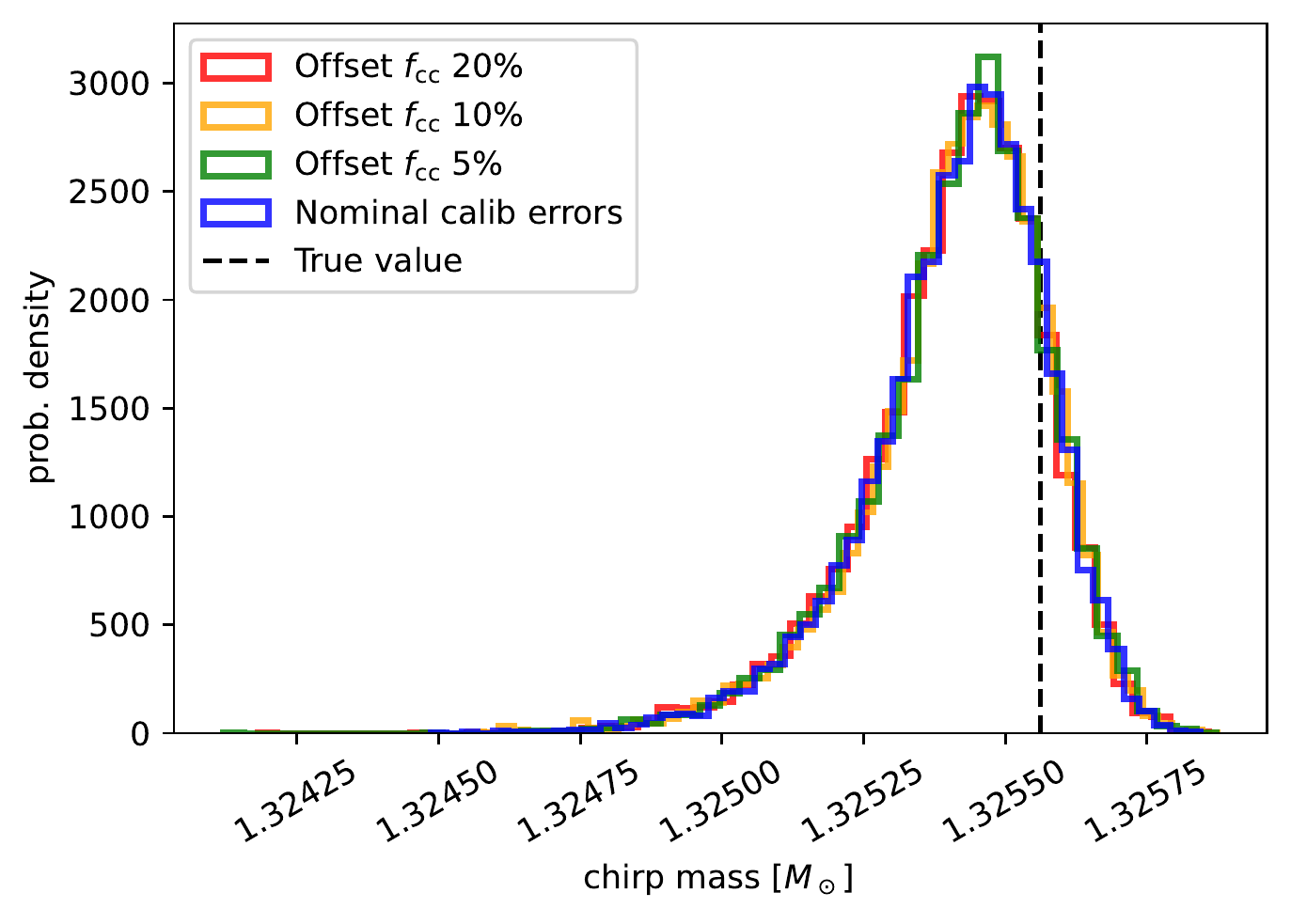}
  \label{fig:sub2}
\end{subfigure}
\begin{subfigure}{.5\textwidth}
  \centering
  \includegraphics[width=\linewidth]{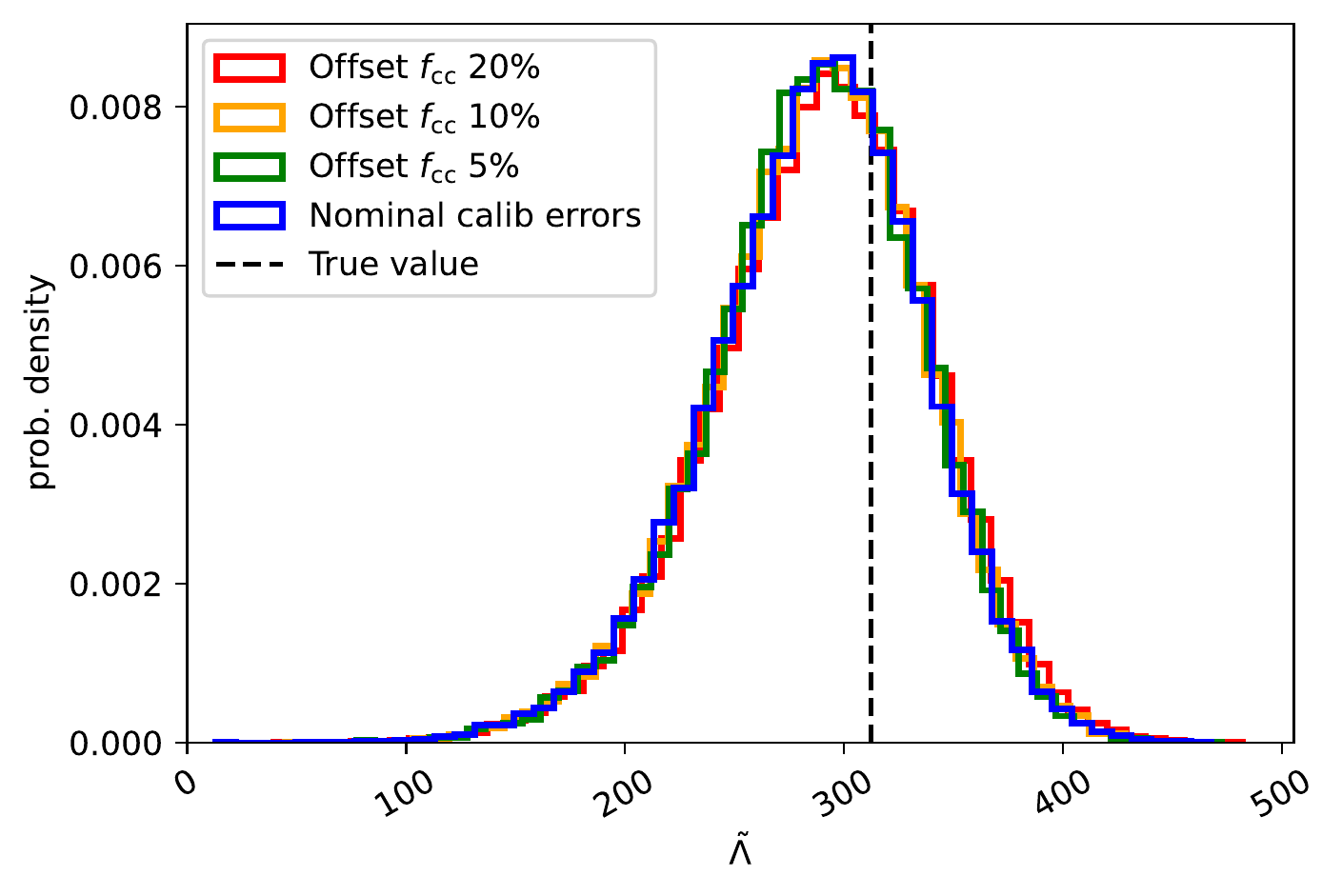}
  \label{fig:sub1}
\end{subfigure}%
\begin{subfigure}{.5\textwidth}
  \centering
  \includegraphics[width=\linewidth]{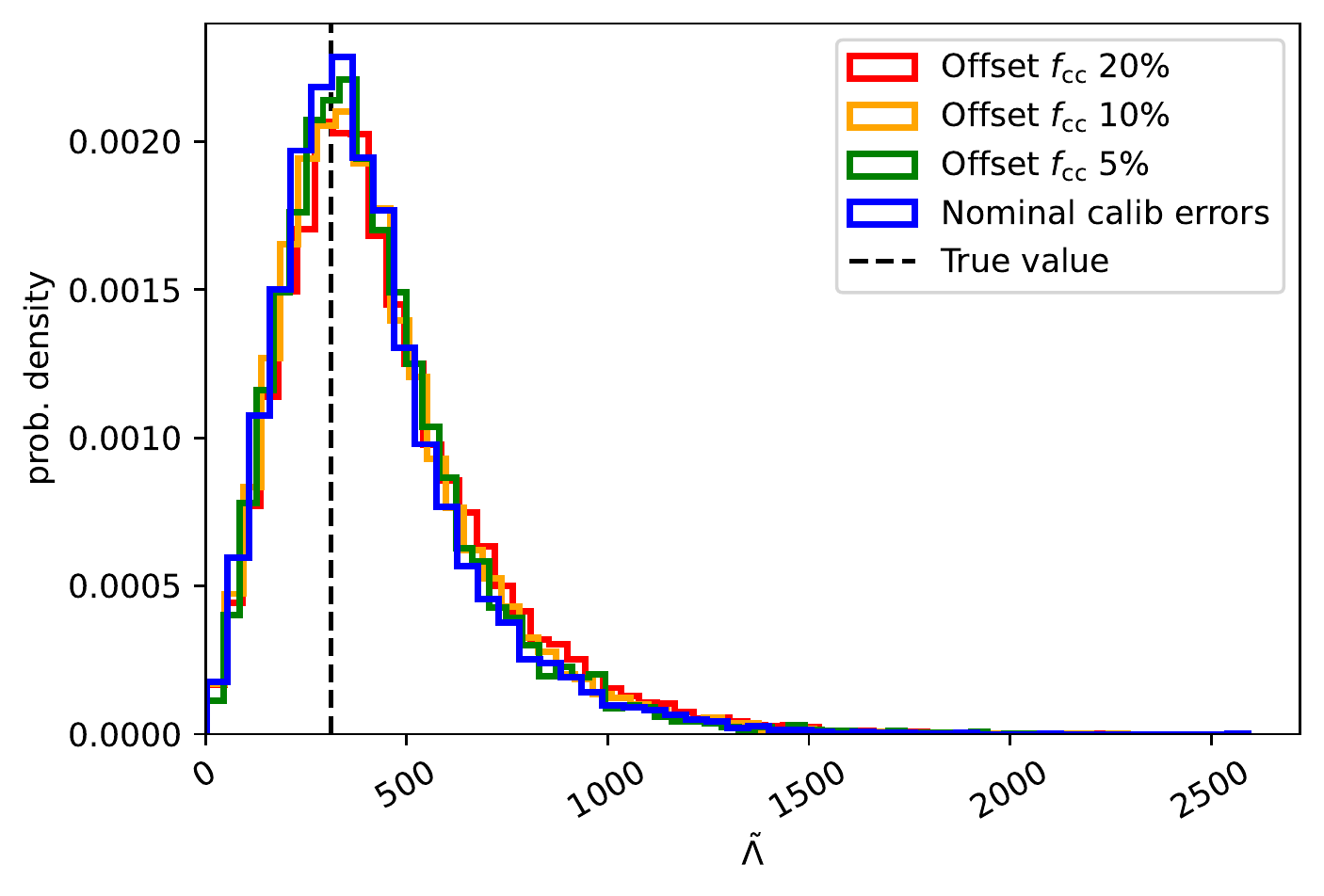}
  \label{fig:sub2}
\end{subfigure}
\caption{\label{fig:fcc_results} Posteriors for the chirp mass $\mathcal{M}$ (top) and tidal deformability $\tilde \Lambda$ (bottom) from the simulated BNS signal with an SNR of 56 (left) and the same signal with an SNR of 16 (right).  These plots show the posteriors for the situation with nominal calibration errors applied to the signal and with calibration errors computed from a 5\%, 10\% and 20\% offset of $f_{\rm cc}$ from its nominal value.  The black dashed line shows the true parameter value.  The results indicate no significant difference between the situation with nominal calibration errors and the situation with a 5\%, 10\% or 20\% offset in the $f_{\rm cc}$ value, which suggests that errors induced in the calibration by a lack of compensation for the TDCF $f_{\rm cc}(t)$ would not significantly impact the results of source parameter estimation for this type of BNS signal.  The slight apparent measurement bias in $\mathcal{M}$ is due to marginalization over mass ratio $m_2 / m_1$ to obtain the one-dimensional posterior distribution.}
\end{figure*}

\begin{figure*}
\centering
\begin{subfigure}{.5\textwidth}
  \centering
  \includegraphics[width=\linewidth]{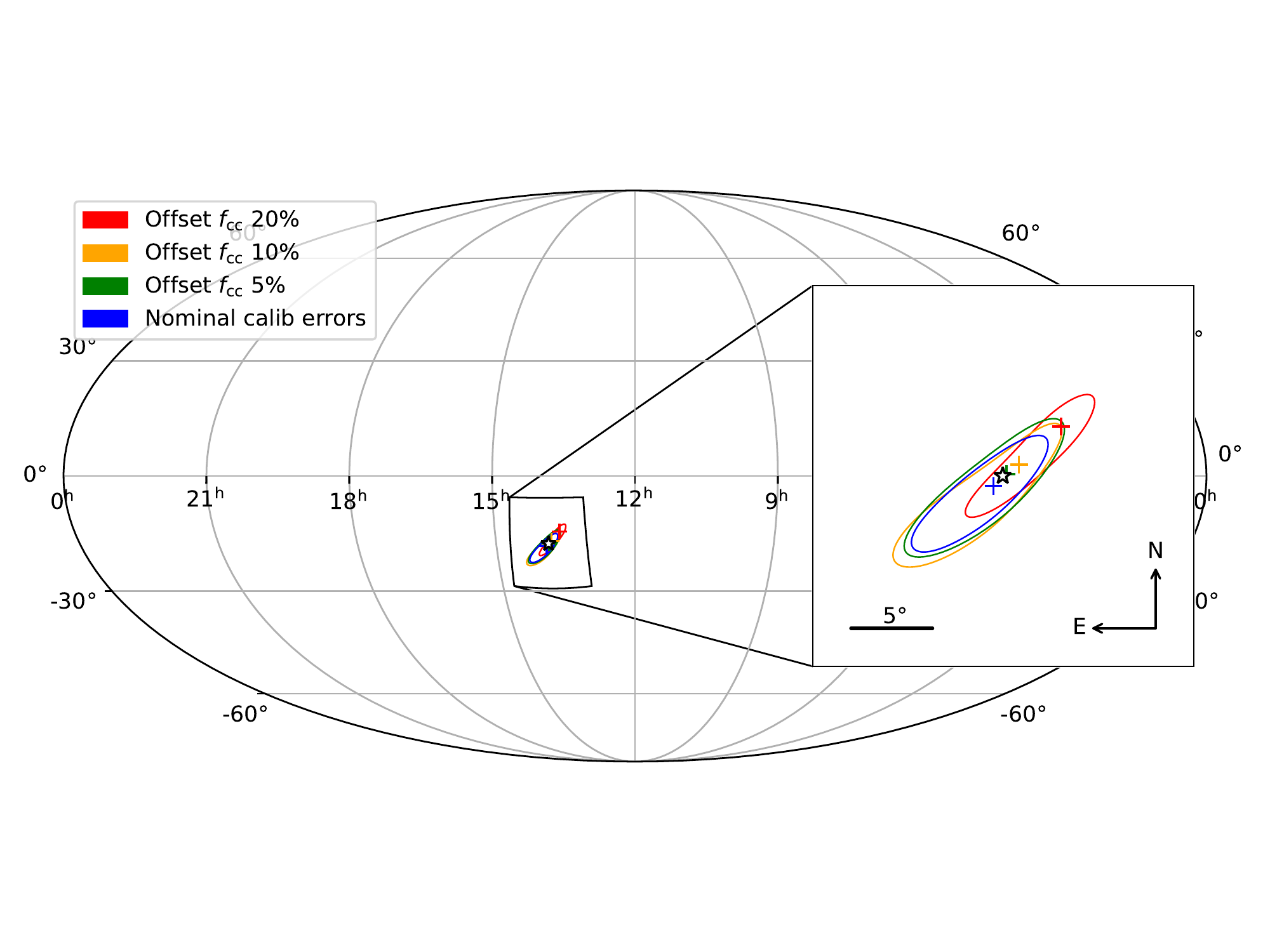}
  \label{fig:sub1}
\end{subfigure}%
\begin{subfigure}{.5\textwidth}
  \centering
  \includegraphics[width=\linewidth]{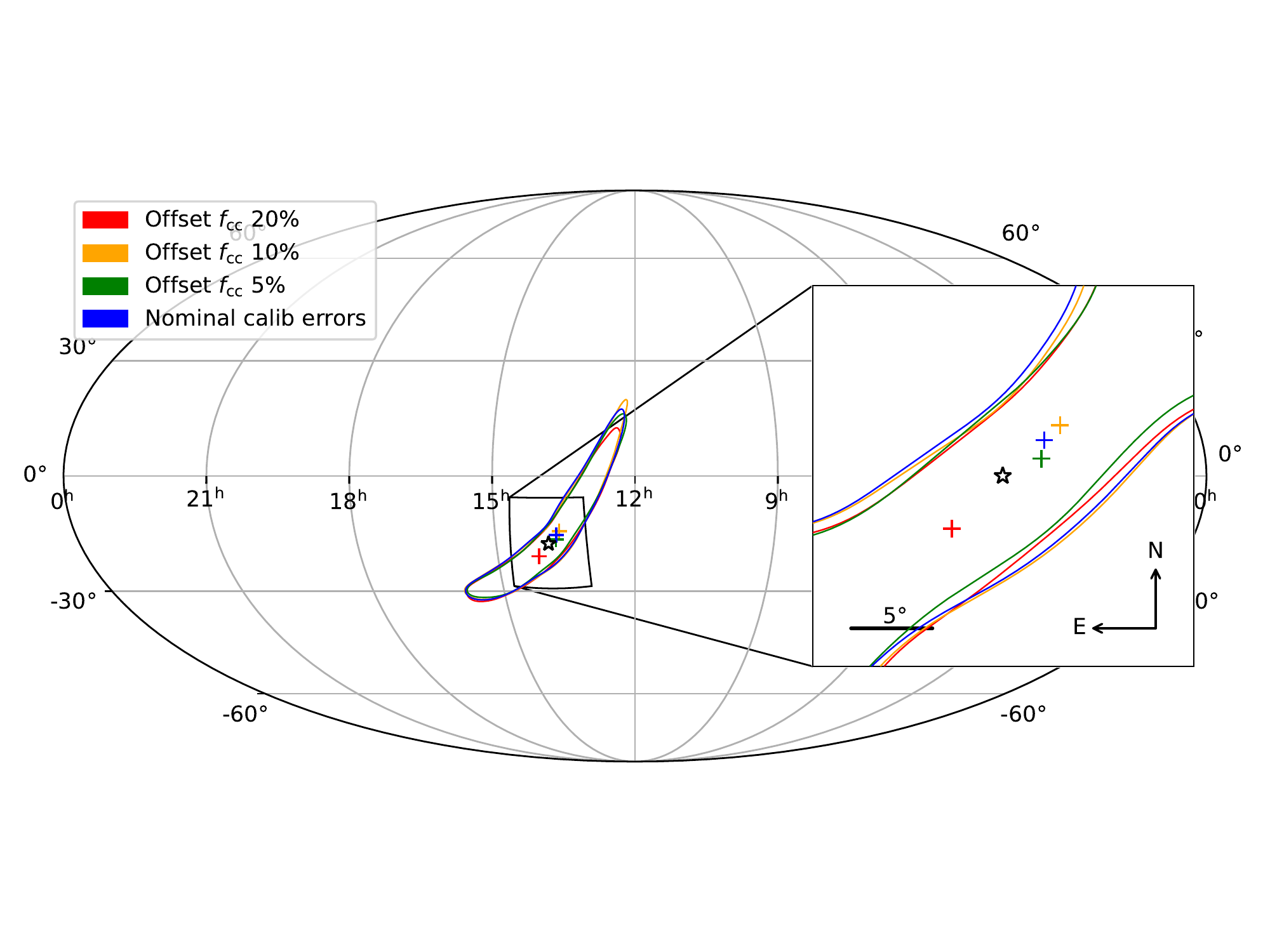}
  \label{fig:sub2}
\end{subfigure}
\caption{\label{fig:skymaps_fcc} Sky maps generated using the \texttt{ligo.skymap} software package~\cite{PhysRevD.93.024013} for the simulated BNS signal with an SNR of 56 (left) and the same signal with an SNR of 16 (right).  Each contour represents the 90\% credible interval for the sky location for different offsets of the cavity pole frequency TDCF $f_{\rm cc}$.  The star represents the true sky position of the simulated signal.  The $+$ indicates the maximum posterior sky position value for each corresponding data set.}  
\end{figure*}

\begin{table*}
\centering
\begin{tabular}{c | c c | c c | c c }
\hline
& \multicolumn{2}{c |}{90\% area (sq. deg.)} & \multicolumn{2}{c |}{Area w/ true loc. (sq. deg.)} & \multicolumn{2}{c}{Prob. w/ true loc. (\%)} \\ [0.5ex]
\hline
& High SNR & Low SNR & High SNR & Low SNR & High SNR & Low SNR \\
\hline\hline
Nominal calib. errors & 29.2 & 435 & 1.72 & 9.51 & 17 & 7 \\
Offset $f_{\rm cc}$ 5\% & 31.1 & 359 & 4.53 & 3.57 & 32 & 4 \\
Offset $f_{\rm cc}$ 10\% & 31.4 & 408 & 12.3 & 19.6 & 61 & 13 \\
Offset $f_{\rm cc}$ 20\% & 31.9 & 381 & 27.9 & 1.60 & 87 & 2 \\
\hline
\end{tabular}
\caption{\label{tab:fcc_results} Summary of sky map statistics generated by the \texttt{ligo.skymap} software package~\cite{PhysRevD.93.024013} for the simulated BNS signal with nominal calibration errors applied as well as calibration errors including offsets of the $f_{\rm cc}$ parameter.  The first two columns show the area of the sky map in units of square degrees enclosed by the measured 90\% credible interval for both the higher-SNR and lower-SNR signal.  The third and fourth columns show the area of the smallest credible region that includes the true sky location for the higher and lower-SNR signals.  The fifth and sixth columns show the smallest credible region percentage that includes the true sky location. The higher-SNR signal shows an increase in the area of the 90\% credible region as well as increase in the area and probability region containing the true signal as the calibration error increases.  There is not a clear pattern that emerges for the lower-SNR signal.}
\end{table*}

\begin{figure*}
\centering
\begin{subfigure}{.5\textwidth}
  \centering
  \includegraphics[width=\linewidth]{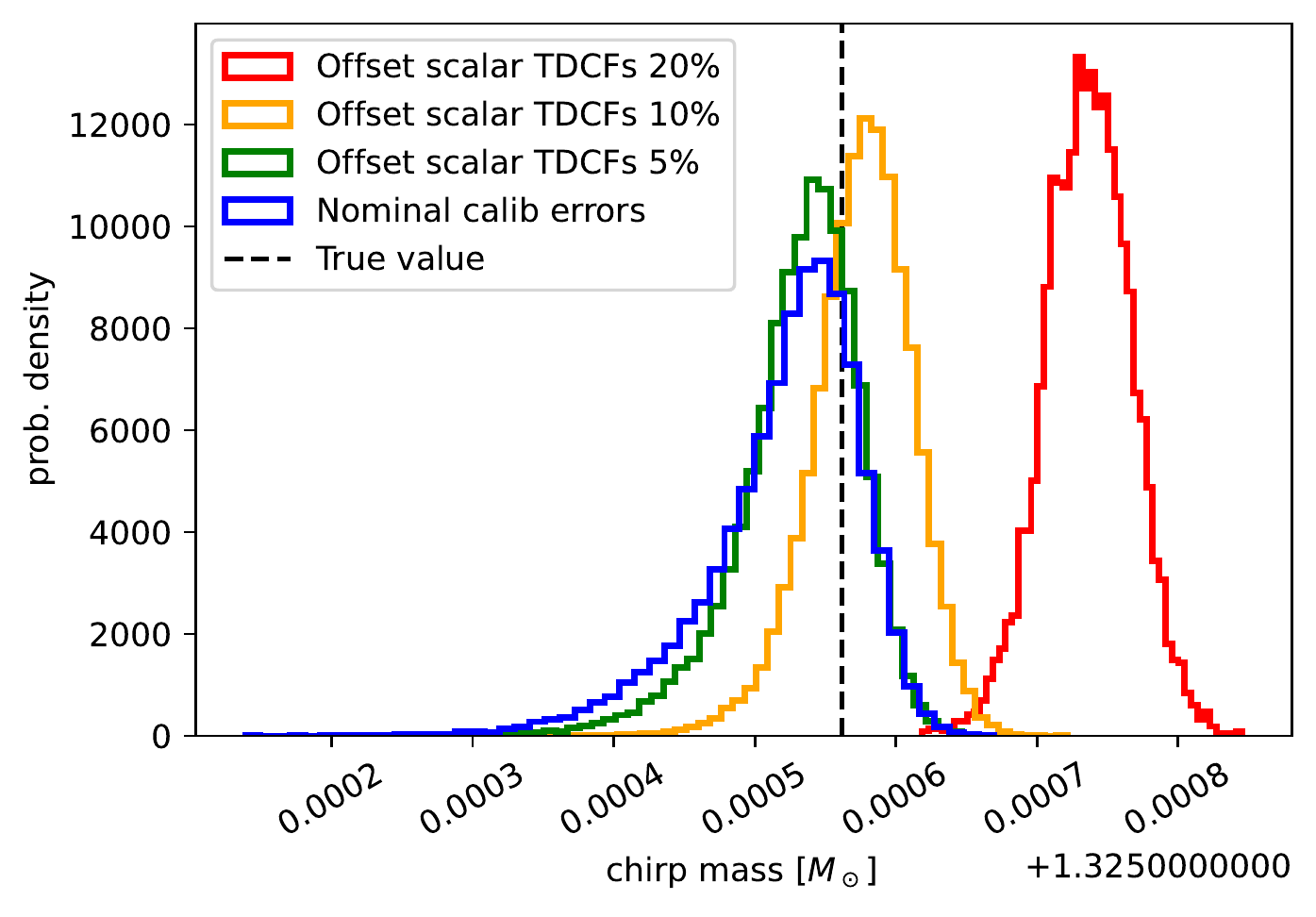}
  \label{fig:sub1}
\end{subfigure}%
\begin{subfigure}{.5\textwidth}
  \centering
  \includegraphics[width=\linewidth]{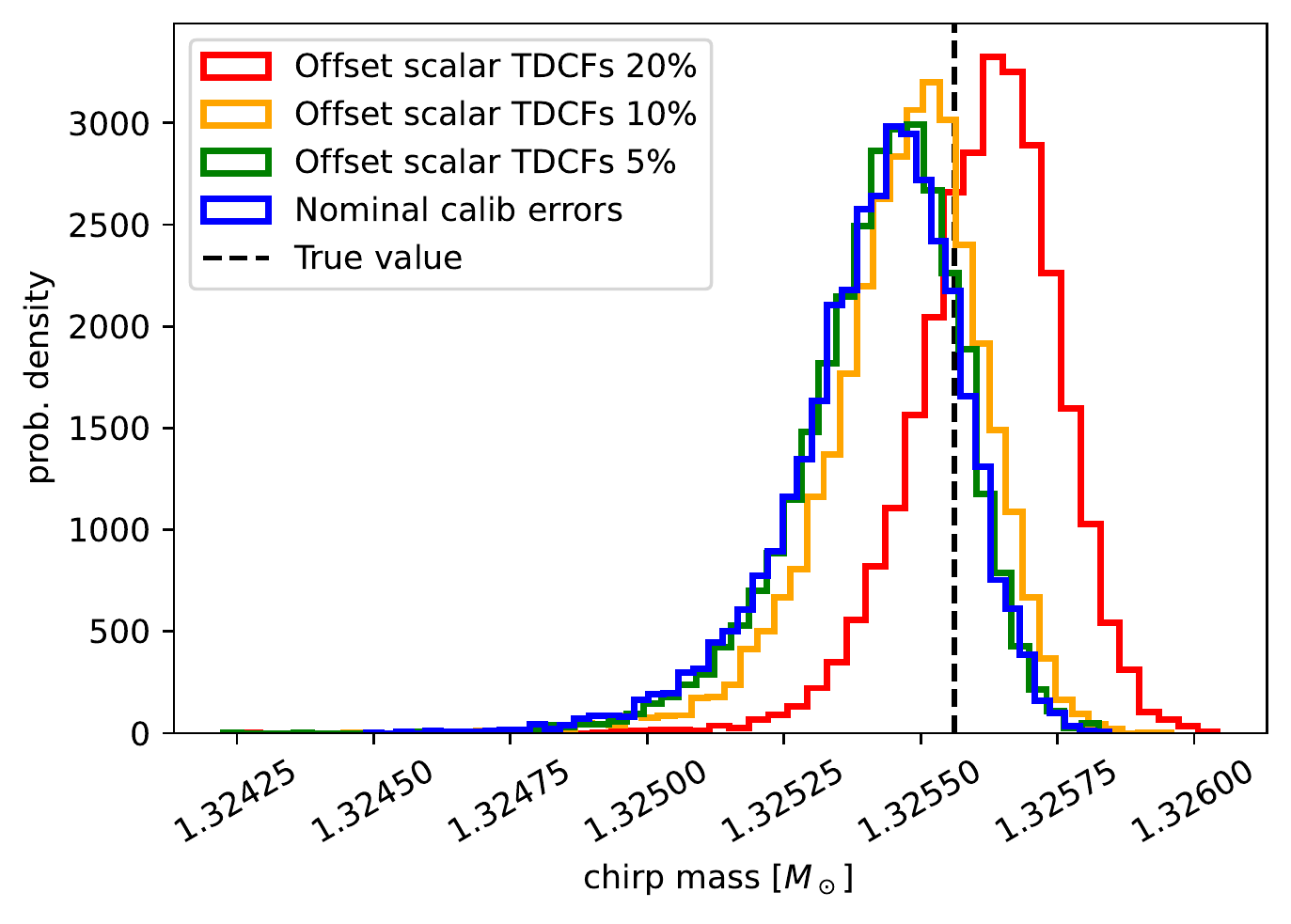}
  \label{fig:sub2}
\end{subfigure}
\begin{subfigure}{.5\textwidth}
  \centering
  \includegraphics[width=\linewidth]{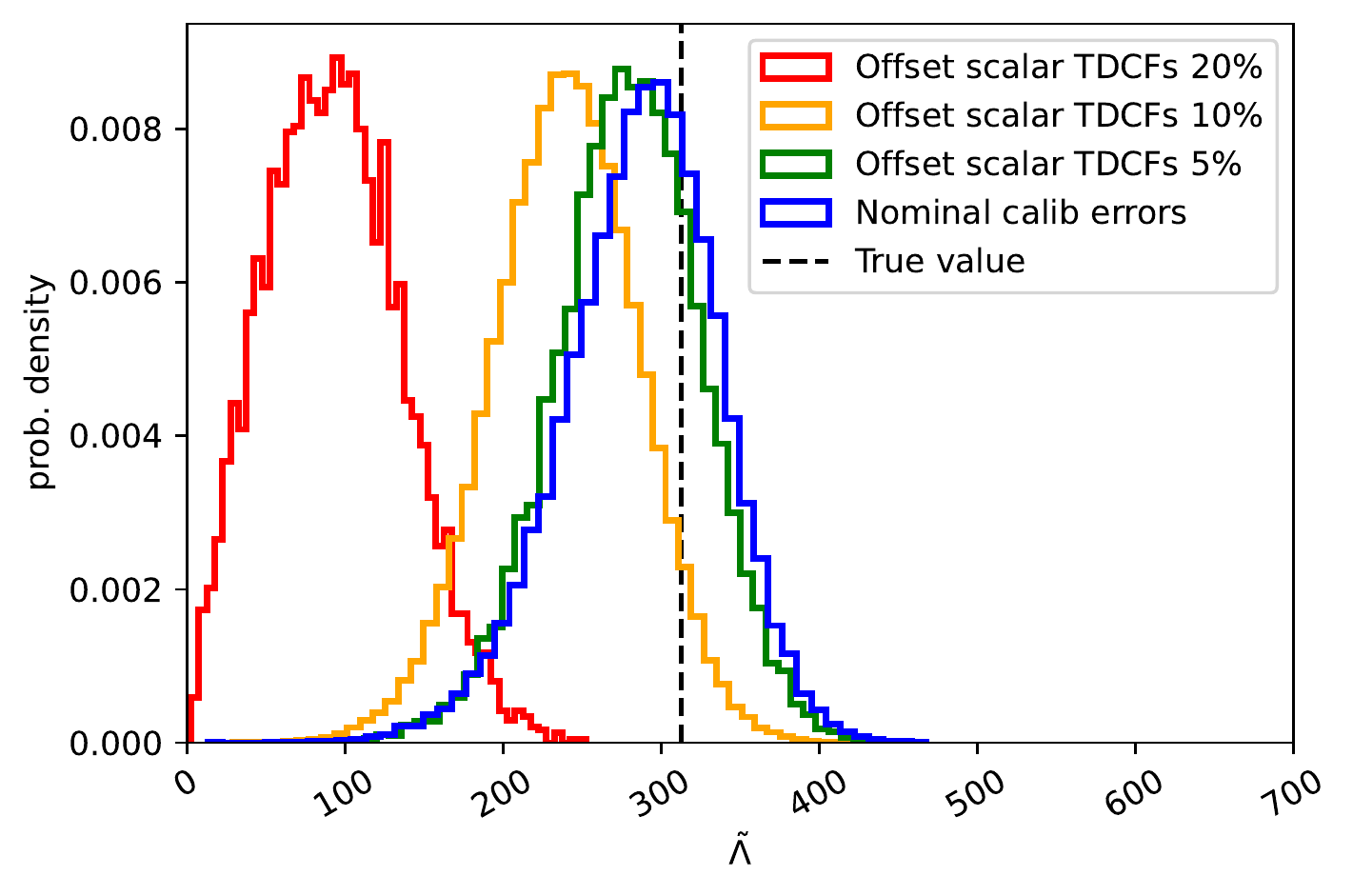}
  \label{fig:sub1}
\end{subfigure}%
\begin{subfigure}{.5\textwidth}
  \centering
  \includegraphics[width=\linewidth]{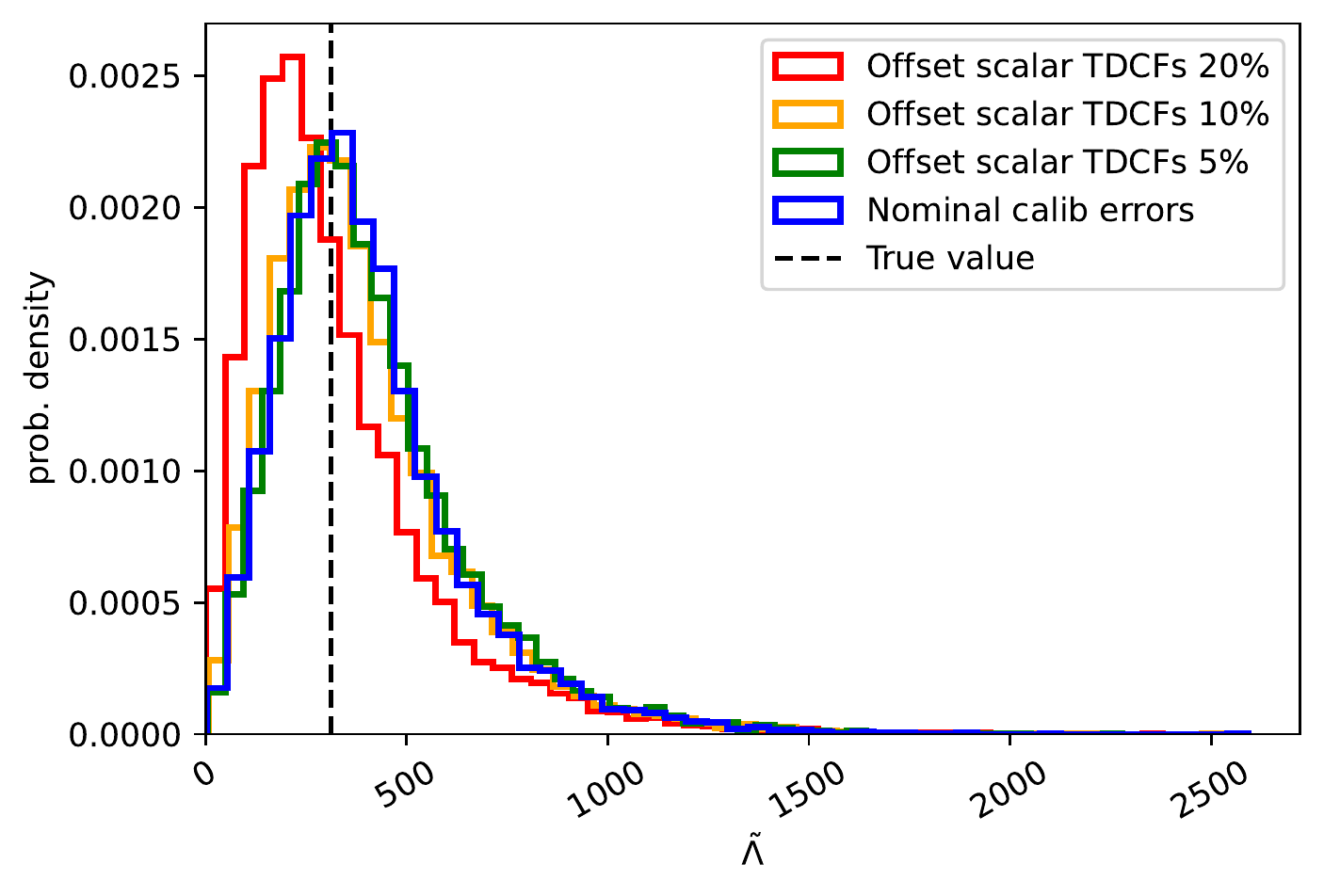}
  \label{fig:sub2}
\end{subfigure}
\caption{\label{fig:alltdcf_results} Posteriors for the chirp mass $\mathcal{M}$ (top) and tidal deformability $\tilde \Lambda$ (bottom) from the simulated BNS signal with an SNR of 56 (left) and the same signal with an SNR of 16 (right).  These plots show the posteriors for the situation with nominal calibration errors applied to the signal and with calibration errors computed from a 5\%, 10\% and a 20\% offset of the TDCF multipliers ($\kappa_{\rm c}$, $\kappa_{\rm T}$, $\kappa_{\rm P}$, and $\kappa_{\rm U}$) from their nominal values.  The black dashed line indicates the true parameter value.  The results show a measurable difference in both $\mathcal{M}$ and $\tilde \Lambda$ for both the high and low SNR signals when the TDCF multipliers are offset by 20\%.  When the TDCF multipliers are offset by 10\% there is a measurable bias in the $\mathcal{M}$ and $\tilde \Lambda$ for the higher-SNR signal, but there is no significant bias introduced for the lower-SNR signal.  The slight apparent measurement bias in $\mathcal{M}$ for the nominal calibration errors and 5\% offset in TDCF multpliers is due to marginalization over mass ratio $m_2 / m_1$ to obtain the one-dimensional posterior distribution.}
\end{figure*}

\begin{figure*}
\centering
\begin{subfigure}{.5\textwidth}
  \centering
  \includegraphics[width=\linewidth]{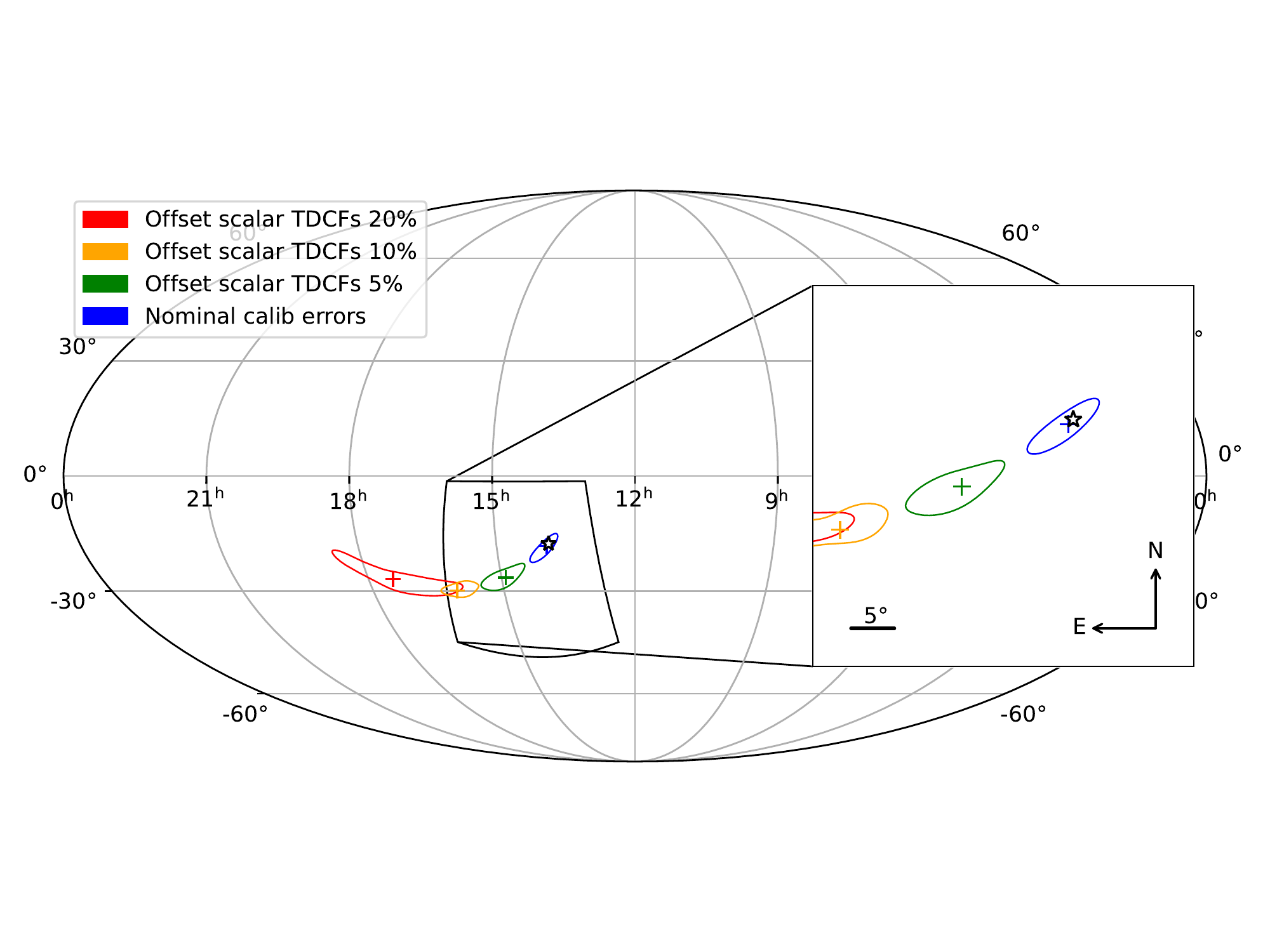}
  \label{fig:sub1}
\end{subfigure}%
\begin{subfigure}{.5\textwidth}
  \centering
  \includegraphics[width=\linewidth]{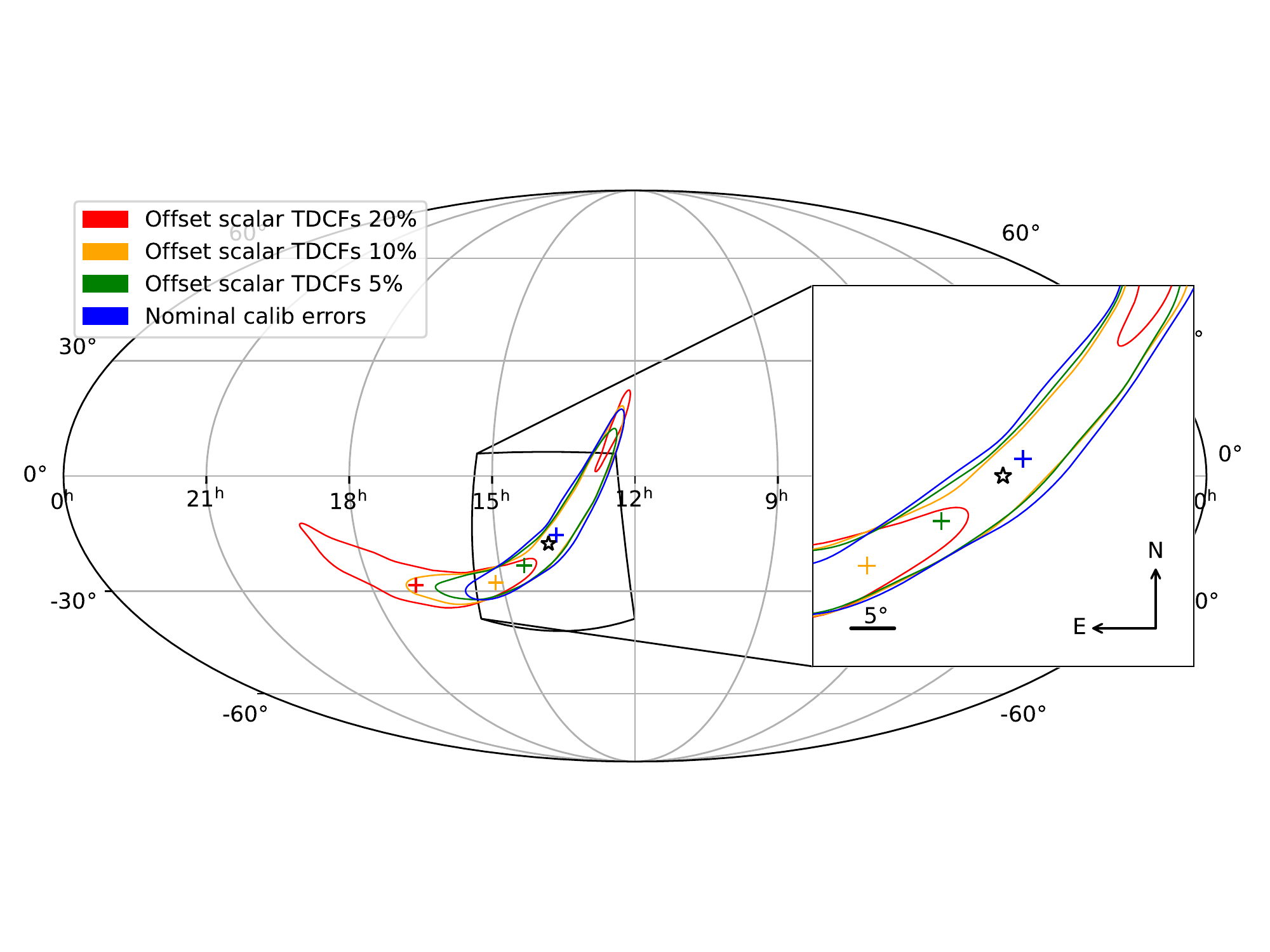}
  \label{fig:sub2}
\end{subfigure}
\caption{\label{fig:skymaps_alltdcf} Sky maps generated using the \texttt{ligo.skymap} software package \cite{PhysRevD.93.024013} for the simulated BNS signal with an SNR of 56 (left) and the same signal with an SNR of 16 (right).  Each contour represents the 90\% credible interval for sky location for different offsets of the TDCF multipliers ($\kappa_{\rm c}$, $\kappa_{\rm T}$, $\kappa_{\rm P}$, and $\kappa_{\rm U}$).  The star represents the true sky position of the simulated signal.  The $+$ indicates the maximum posterior sky position value for each corresponding data set.}  
\end{figure*}

\begin{figure}[h!]
\centering
\includegraphics[width=\columnwidth]{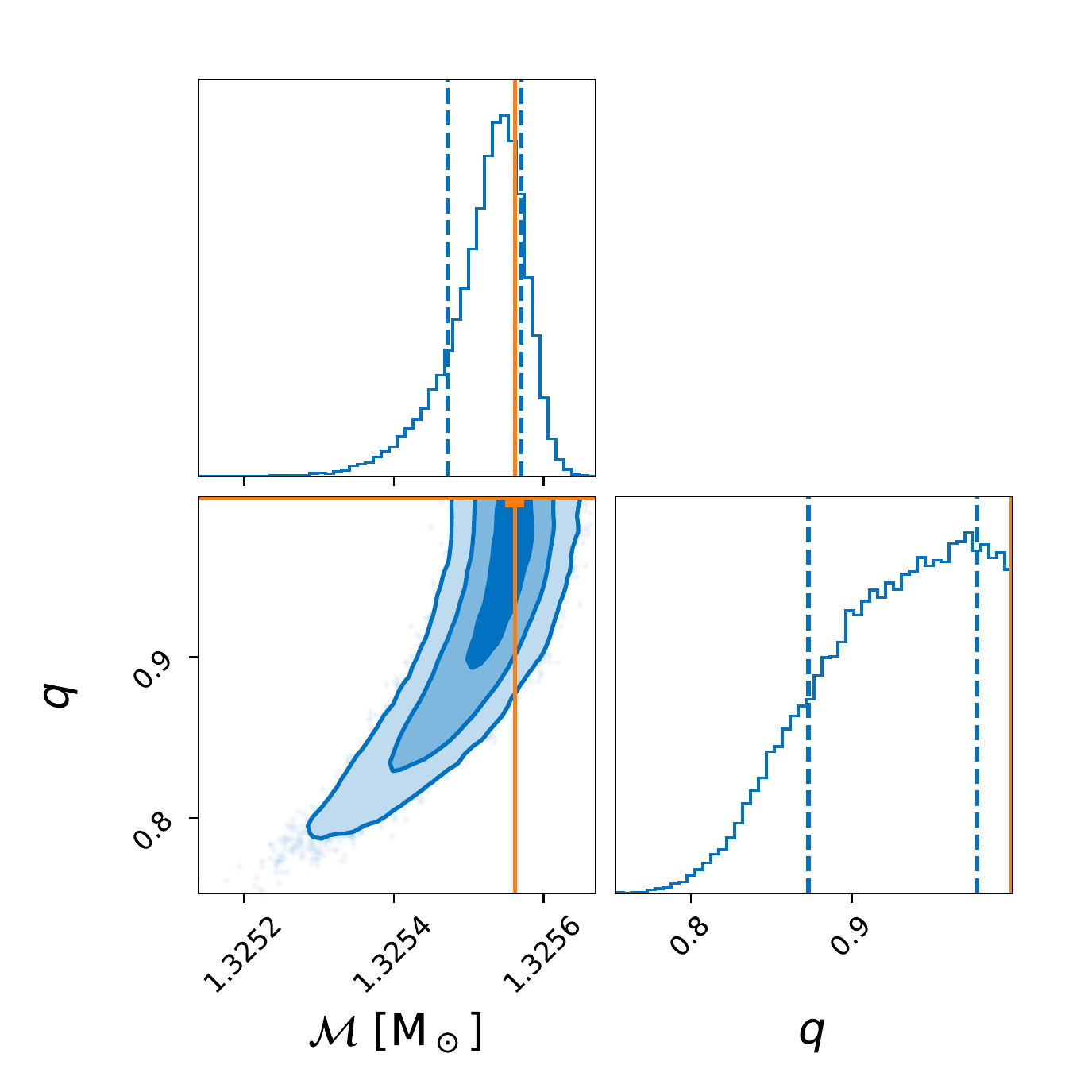}
\caption{\label{fig:mc_q} Corner plot for the chirp mass $\mathcal{M}$ and mass ratio $q$ parameters for the data set where nominal calibration errors were applied to a BNS injection with an SNR of 56.  This corner plot illustrates how the one-dimensional $\mathcal{M}$ posterior distribution contains an apparent bias due to the marginalization over $q$.}
\end{figure}

While we jointly inferred in all 11 waveform parameters, we highlight here the estimation of two source parameters: chirp mass $\mathcal{M}$ and the combined tidal deformability parameter $\tilde \Lambda$.  
These parameters are both given by linear combinations of the binary component parameters, 
\begin{eqnarray}
\mathcal{M} &=& \frac{(m_1 m_2)^{3/5}}{(m_1+m_2)^{1/5}} \\
\nonumber
\tilde \Lambda &=& \frac{8}{13}\left[ \left(1+7\eta - 31 \eta^2\right)\left(\Lambda_1 + \Lambda_2\right) \right .+ \\
&& \left . \sqrt{1-4\eta} \left(1+9\eta-11\eta^2\right) \left(\Lambda_1 - \Lambda_2\right)\right] \ ,
\end{eqnarray}
where $\eta = m_1 m_2 / (m_1+m_2)^2$ is the symmetric mass ratio.
Chirp mass is one of the most precisely measured parameters encoded in the gravitational waves of binary coalescence events.  
The tidal deformability parameter $\tilde \Lambda$ is the most precisely measured parameter related to neutron star matter deformability and is known to be an effect that appears at high frequencies~\cite{PhysRevD.77.021502,PhysRevD.81.123016,PhysRevLett.112.101101,Wade:2014vqa}.
For this reason, we investigated whether systematic errors introduced by not compensating for temporal variations of $f_{\rm cc}$, which is also a parameter that will impact the data at higher frequencies (around a few hundred Hz and above), would impact the estimation of $\tilde \Lambda$.

In addition to highlighting the posteriors for $\mathcal{M}$ and $\tilde \Lambda$, we produced sky maps from the posterior samples using the \texttt{ligo.skymap} software package \cite{PhysRevD.93.024013} showing the 90\% credible intervals for the sky localization of each signal.
As mentioned above, we investigated whether relatively maximal offsets between the H1 and L1 detector would impact sky localization, since this is largely determined from the relative timing of the signal between two or more detectors.
The sky maps are also accompanied by two relevant statistics. 
The first statistic is related to the precision of sky localization and is a measure of the area in square degrees of the 90\% credible region.
The second statistic is related to the accuracy of the sky localization and is constructed by searching for the smallest credible region that would contain the true sky location.
Each of these statistics are discussed in more detail in Ref.~\cite{PhysRevD.93.024013}.

Results are shown in Figs.~\ref{fig:fcc_results} -- \ref{fig:skymaps_alltdcf} and Tabs.~\ref{tab:fcc_results} and~\ref{tab:alltdcf_results}.
The results shown in Fig.~\ref{fig:fcc_results} indicate that systematic calibration errors introduced by not compensating for temporal variations in the coupled cavity pole frequency parameter $f_{\rm cc}$ have a no measurable impact on the estimation of $\mathcal{M}$ and $\tilde \Lambda$.
This is true for both the lower-SNR signal (SNR of 16) and the higher-SNR signal (SNR of 56).
The slight apparent measurement bias in $\mathcal{M}$ is due to marginalization over mass ratio $q = m_2 / m_1$ to obtain the one-dimensional posterior distribution.
Since we assume $m_1 > m_2$, the injected value of $q$ is very close to its maximum possible value of 1.
Fig.~\ref{fig:mc_q} is a corner plot in $\mathcal{M}$ and $q$, illustrating how the marginalization over $q$ will skew the peak of a one-dimensional posterior distribution for $\mathcal{M}$.

Sky localization of the higher-SNR signal is impacted slightly by reasonably low uncompensated variations in $f_{\rm cc}$, as shown in Fig.~\ref{fig:skymaps_fcc} and Tab.~\ref{tab:fcc_results}.
Increasing calibration errors do lead to an increase in the area enclosed by the 90\% credible interval for the higher-SNR signal.
Additionally, there is a bias introduced into the sky localization for the higher-SNR signal that increases as the size of the calibration error increases.  
Tab.~\ref{tab:fcc_results} shows the area enclosed by the smallest credible interval containing the true sky location as well as the probability that corresponds to this credible interval.
The lower-SNR signal, however, does not show a clear trend in the uncertainty of the sky localization, quantified by the area enclosed by the 90\% credible interval, or the bias introduced into the sky localization, quantified by the credible interval containing the true sky location, as calibration error is increased.  

When we focused on systematic calibration errors induced by not compensating for the TDCF multipliers, we did see biases enter the source parameter estimation results.
We observed noticeable changes in the measurability of source parameters when the TDCF multipliers were offset from their nominal values by 20\% for both the higher and lower-SNR signals, as shown in Fig.~\ref{fig:alltdcf_results}.
When the TDCF multipliers were offset by 10\% from their nominal values there was a bias in the recovered source parameters for the higher-SNR signal only.
For the 10\% offset, the lower-SNR signal was still dominated by statistical error.
No noticeable change in measurable parameters was seen for a 5\% offset in the TDCF multipliers.

The most impactful consequence of calibration errors induced from the lack of compensation for the TDCF multipliers comes through in the sky localization of each signal, as shown in Fig.~\ref{fig:skymaps_alltdcf} and Tab.~\ref{tab:alltdcf_results}. 
For both signals, the area enclosed by the 90\% credible region increases as the calibration error increases, with a noticeable increase for even the 5\% offset of scalar TDCFs.
The lower SNR-signal also shows a steady increase to the bias of the sky localization as the calibration error increases.
By the time a 10\% offset to all scalar TDCFs is introduced, the higher-SNR signal sky localization is only found in a credible region larger than 99\%, which demonstrates that even a 10\% offset to the scalar TDCFs is devastating to our ability to locate the signal on the sky if such an offset to scalar TDCFs were left uncompensated.
It is critical to note, however, that the released calibrated data has always compensated for scalar TDCFs and therefore has not included such errors.

\begin{table*}
\centering
\begin{tabular}{c | c c | c c | c c }
\hline
& \multicolumn{2}{c |}{90\% area (sq. deg.)} & \multicolumn{2}{c |}{Area w/ true loc. (sq. deg.)} & \multicolumn{2}{c}{Prob. w/ true loc. (\%)} \\ [0.5ex]
\hline
& High SNR & Low SNR & High SNR & Low SNR & High SNR & Low SNR \\
\hline\hline
Nominal calib. errors & 29.2 & 435 & 1.72 & 9.51 & 17 & 7 \\
Offset scalar TDCFs 5\% & 68.2 & 364 & 69.7 & 21.3 & 90 & 16 \\
Offset scalar TDCFs 10\% & 56.2 & 389 & $>1000$ & 79.3 & $>99$ & 36 \\
Offset scalar TDCFs 20\% & 114.7 & 587 & $>1000$ & 857 & $>99$ & 95 \\
\hline
\end{tabular}
\caption{\label{tab:alltdcf_results} Summary of sky map statistics generated by the \texttt{ligo.skymap} software package~\cite{PhysRevD.93.024013} for the simulated BNS signal with nominal calibration errors applied as well as calibration errors including offsets of the TDCF multipliers ($\kappa_{\rm c}$, $\kappa_{\rm T}$, $\kappa_{\rm P}$, and $\kappa_{\rm U}$).  The first two columns show the area of the sky map in units of square degrees enclosed by the measured 90\% credible region for both the higher-SNR and lower-SNR signal.  The third and fourth columns show the area of the smallest credible region that includes the true sky location for the higher and lower-SNR signals.  The fifth and sixth columns show the smallest credible region percentage that includes the true sky location. The higher and lower-SNR signals both shows an increase in the area of the 90\% credible region as the calibration error increases.  The lower-SNR signal also shows a steady increase in the area and probability region containing the true signal as the calibration error increases.  However, the higher-SNR signal caps out at more than 99\% credible region containing the true sky location by the time we reach an offset of 10\% in the scalar TDCFs.  Therefore, the change between the 10\% and 20\% scalar TDCF offset results are negligible.  By the time an uncompensated 10\% offset in the scalar TDCFs is reached, the sky localization is severely impacted.}
\end{table*}

In summary, we found that calibration errors induced by not correcting for the TDCF $f_{\rm cc}$ is only impactful through the sky localization. 
In particular, for louder signals we would be able to detect a broadening of the sky location credible regions and a bias introduced in the sky localization if variations of the TDCF $f_{\rm cc}$ were left uncompensated.  
Not compensating for the TDCF multipliers ($\kappa_{\rm c}$, $\kappa_{\rm T}$, $\kappa_{\rm P}$, $\kappa_{\rm U}$) can lead to significant biases in the estimation of source parameters, especially for louder signals when offsets of the scalar TDCFs reach 10\% of their nominal values.
Other studies \cite{Payne:2020myg, Vitale:2020gvb, Huang:2022rdg} have investigated these questions in more detail, including in their studies the uncertainty on the calibration systematic error, and work is ongoing to better understand the impact of calibration errors outside of source parameter estimation for compact binary events.

Our limited study here aimed to investigate specifically how not compensating for TDCF filters, such as $f_{\rm cc}$, in the calibration procedure could impact a subset of astrophysical results, since we did know that not compensating for TDCF filters leaves a measurable systematic error in the calibrated strain data (see Sec.~\ref{sec:calibrationAccuracy}).
We found that compensation for the TDCF $f_{\rm cc}$ is important when considering sky localization.
We also demonstrated that correcting for more general time dependence of the calibration model through the TDCF multipliers can have a significant impact on source parameter estimation including the sky localization of a signal.

\section{Conclusion}
\label{sec:conclusion}
Temporally varying filters in the sensing and actuation functions have been a source of systematic error in Advanced LIGO's calibrated data when left uncompensated, as shown in Sec.~\ref{sec:calibrationAccuracy}.
In order to correct these systematic errors, we developed adaptive filtering algorithms in the \texttt{gstlal} calibration pipeline.  
Compensating for the time dependence of the coupled cavity pole $f_{\rm cc}$, as was done in an offline version of the O2 calibration and throughout O3, removes most of the systematic error above $\sim$100 Hz, with negligible errors remaining at the $f^{\rm pc}_2 \sim 400$~Hz Pcal line.

In order to remove the remaining systematic error at the $f^{\rm pc}_1 \sim 10-40$~Hz Pcal line, the calibration needs to additionally compensate for temporal variations in the low-frequency regime of the sensing function.
The low-frequency variations of the sensing function were not sufficiently well modeled during previous observing runs to compensate appropriately, which left a lingering systematic error in this frequency region.  
While compensating for the calibration model parameters $f_s$, $Q$ and time delays $\tau_i$ can improve the calibration accuracy exactly at the Pcal lines used to compute these corrections, these compensations did not lead to improved broadband calibration accuracy in most situations, especially during O3 (Fig.~\ref{fig:pcalBroadbandManual}).


Further improvement in the calibration accuracy may be possible by improving the estimates of the TDCFs.  
Decreasing the separation of the calibration line frequencies $f_{\rm ctrl}$, $f_{\rm T}$, and $f^{\rm pc}_1$ has been shown to improve the estimates of the TDCFs.
An exact algebraic solution for all the TDCFs, although challenging, is likely the next best improvement to the calibration accuracy related to time dependence of the calibration models~\cite{VietsDissertation}.

We demonstrated that compensating for both TDCF multipliers and filters improves the overall accuracy of the calibrated strain data.
An open question from here is how much this improved accuracy will lead to improved results in astrophysical analyses that flow from the calibrated strain data.
We investigated a narrow version of this question by studying the impact of simulated systematic calibration errors that would be caused by not compensating for different levels of temporal variations in the calibration model, isolating TDCF filters from TDCF multipliers.
For this study, we looked at a simulated BNS signal with two different SNRs.
Systematic calibration errors caused by not compensating for TDCF filters could be impactful when determining the sky localization of louder events.
Additionally, systematic calibration errors caused by not compensating for TDCF multipliers can lead to significant biases in the source parameter estimation, including sky localization. 
It's important to note that all released LIGO strain data contains the appropriate compensation for TDCF multipliers and filters, using methods such as those described in this work. 

While this work was ongoing, other researchers~\cite{Payne:2020myg, Vitale:2020gvb} developed a more sophisticated framework to build the calibration systematic error and its associated uncertainty into source parameter estimation algorithms directly.
In the future, the type of infrastructure developed by Payne et al. and Vitale et al. can be used to investigate specific calibration error scenarios to help inform future development related to compensating for temporal variations in the calibration models.

From this work alone we can conclude that ongoing compensation for the scalar TDCFs is critical in the final calibrated strain data products.  
Additionally, compensation for TDCF filters should be included whenever possible to ensure accurate sky localization results are obtained, especially as louder signals become more commonplace with improving sensitivity of the LIGO detectors.

\begin{acknowledgments}
The authors would like to thank Lilli Sun for her helpful suggestions and careful review of this paper.
The authors would like to thank the members of the LIGO calibration team, including Joe Betzwieser, Dripta Bhattacharjee, Jenne Driggers, Evan Goetz, Sudarshan Karki, Jeff Kissel, Antonios Kontos, Greg Mendell, Timesh Mistry, Ethan Payne, Jamie Rollins, Rick Savage, Lilli Sun, and Alan Weinstein.
The authors were supported by National Science Foundation grants PHY-1607178, PHY-1607585, PHY-1506360, and PHY-1847350.  
LIGO was constructed by the California Institute of Technology and Massachusetts Institute of Technology with funding from the United States National Science Foundation (NSF), and operates under cooperative agreement PHY–1764464. 
The authors are grateful for computational resources provided by the LIGO Laboratory and supported by National Science Foundation Grants PHY-0757058 and PHY-0823459.
This material is based upon work supported by NSF's LIGO Laboratory which is a major facility fully funded by the National Science Foundation.
Advanced LIGO was built under award PHY–0823459. 
The authors gratefully acknowledge the support of the United States NSF for the construction and operation of the LIGO Laboratory and Advanced LIGO as well as the Science and Technology Facilities Council (STFC) of the United Kingdom, the Max-Planck-Society (MPS), and the State of Niedersachsen/Germany for support of the construction of Advanced LIGO and construction and operation of the GEO600 detector. 
Additional support for Advanced LIGO was provided by the Australian Research Council (ARC).
This paper carries LIGO Document Number LIGO-P1800313.

\end{acknowledgments}

\appendix

\section{Time-dependence of Calibration Model Time Delay Parameters}
\label{app:tau}

As mentioned in Secs.~\ref{sec:CalibrationModels} and~\ref{sec:calibrationAccuracy}, compensation for time dependence in the $\tau_i$ does not always lead to improvements in the calibration accuracy.
The calculation of $\tau_i$ can include systematic errors that result in declining accuracy of the calibration when these corrections are applied.  
Estimates of the $\tau_i$ are known to deviate from zero for several reasons:
\begin{itemize}
\item {\it Variable computational time delays in the actuation function.}  Computational delays in the digital portion of $A$ appear to shift on occasion by multiples of the 65-kHz sampling intervals at which operations are done by the computers that run the interferometer control models.  Such shifts manifest themselves as sudden changes in the $\tau_i$, such as those seen in Fig.~\ref{fig:actTiming}.
\item {\it Breakdown of approximations used to estimate the actuation TDCFs.}
The accuracy of these approximations decreases with increasing separation of the calibration line frequencies used to measure the actuation function (see Table~\ref{tab:callines}), and with increasing deviation of the true response function from the nominal response function of the static reference model at those frequencies.  The true response function may deviate from that of the reference model due to changes in the TDCFs or other systematic errors present in the reference model, whether time-dependent or not.  A study reported in \cite{VietsDissertation} showed that systematic errors as large as \SI{40}{\micro\second} could be caused by variations in a single TDCF that are typical during observation.  The other TDCFs also suffered from similar effects.  Additionally, the lack of a complete model for the low-frequency sensing function at H1 can contribute to these systematic errors.
An exact solution for the TDCFs developed in \cite{VietsDissertation} appears to mitigate these problems, allowing for better calibration accuracy achieved by compensating for time dependence in all the TDCFs.  This method may be used during Advanced LIGO's next observing run, O4.
\item {\it Systematic errors in the static reference model.}  In addition to causing the breakdown of approximations noted above, systematic errors in the static reference model can also directly cause erroneous deviations from zero in the $\tau_i$.  This is because the parameters of the reference models are chosen based on all the data collected in the measurement process to produce a model that best matches the data at all frequencies.  Thus, by design, the model may not exactly match the measurements at the calibration line frequencies.  Such errors are generally expected to be small, except in cases where the model does not fit the measurements well, such as can occur at H1 given the incomplete low-frequency sensing function model.
\item {\it Other unknown changes in the frequency dependence of the actuation function.}  It is quite probable that any other changes to the frequency dependence of $A$ are less impactful than the aforementioned effects.  Compensating for such changes would require a new reference model, since we cannot correct for time dependence that has unknown frequency dependence.  Measurements are taken periodically during observing runs to ensure that deviations from the current model are sufficiently small.
\end{itemize}

\bibliographystyle{apsrev4-1}
\bibliography{freq_dep_corrections}

\end{document}